\documentclass[pra,showpacs,twocolumn,aps,superscriptaddress,a4paper,longbibliography]{revtex4-1}
\usepackage{dcolumn,amssymb,amsmath,amsfonts,graphicx,latexsym,color}
\usepackage[utf8]{inputenc}
\usepackage[T1]{fontenc}
\usepackage{epstopdf}

\usepackage{ulem}
\usepackage{epstopdf}
\usepackage{subfigure}
\usepackage{float}
\usepackage{mathrsfs}
\usepackage{bbold}
\usepackage{psfrag}
\usepackage{mathcomp}
\usepackage{verbatim}
\usepackage{multirow}
\usepackage{diagbox}
\usepackage[colorlinks,citecolor=blue]{hyperref}

\usepackage{simplewick}
\begin{document}


\title{Phase transitions in intrinsic magnetic topological insulator with high-frequency pumping}

\author{Fang Qin}
\thanks{They contribute equally to this work. }
\affiliation{Shenzhen Institute for Quantum Science and Engineering and Department of Physics, Southern University of Science and Technology (SUSTech), Shenzhen 518055, China}
\affiliation{CAS Key Laboratory of Quantum Information, University of Science and Technology of China, Chinese Academy of Sciences, Hefei, Anhui 230026, China}
\affiliation{Department of Physics, National University of Singapore, Singapore 117542}

\author{Rui Chen}
\thanks{They contribute equally to this work. }
\affiliation{Shenzhen Institute for Quantum Science and Engineering and Department of Physics, Southern University of Science and Technology (SUSTech), Shenzhen 518055, China}
\affiliation{School of Physics, Southeast University, Nanjing 211189, China}
\affiliation{Department of Physics, The University of Hong Kong, Pokfulam Road, Hong Kong, China}

\author{Hai-Zhou Lu}
\email{Corresponding author: luhz@sustech.edu.cn}
\affiliation{Shenzhen Institute for Quantum Science and Engineering and Department of Physics, Southern University of Science and Technology (SUSTech), Shenzhen 518055, China}
\affiliation{Shenzhen Key Laboratory of Quantum Science and Engineering, Shenzhen 518055, China}

\date{\today }

\begin{abstract}
In this work, we investigate the topological phase transitions in an effective model for a topological thin film with high-frequency pumping. In particular, our results show that the circularly polarized light can break the time-reversal symmetry and induce the quantum anomalous Hall insulator (QAHI) phase. Meanwhile, the bulk magnetic moment can also break the time-reversal symmetry. Therefore, it shows rich phase diagram by tunning the intensity of the light and the thickness of the thin film. Using the parameters fitted by experimental data, we give the topological phase diagram of the Cr-doped Bi$_{2}$Se$_{3}$ thin film, showing that by modulating the strength of the polarized optical field in an experimentally accessible range, there are four different phases: the normal insulator phase, the time-reversal-symmetry-broken quantum spin Hall insulator phase, and two different QAHI phases with opposite Chern numbers. Comparing with the non-doped Bi$_{2}$Se$_{3}$, it is found that the interplay between the light and bulk magnetic moment separates the two different QAHI phases with opposite Chern numbers. The results show that an intrinsic magnetic topological insulator with high-frequency pumping is an ideal platform for further exploring various topological phenomena with a spontaneously broken time-reversal symmetry.
\end{abstract}

\maketitle

\section{Introduction}\label{1}

Topological insulators are band insulators with topologically protected boundary states and insulating bulk states~\cite{Moore2010,Hasan2010,Qi2011,Shen2012,Wang2017,Qin2020,Chen2021prl,Chen2021prr}. A well-known topological insulator paradigm is the quantum Hall insulator (QHI) phase~\cite{ZhangExp2009}.
Distinct from the QHI phase, there also exists a quantized version of the Hall effects in a magnetically doped topological insulator without applying any an external magnetic field.
In an intrinsic magnetic topological insulator~\cite{XueExp2013,CheckelskyExp2014,QAHIexp2014,QAHIexp2015,XueExp2015,YoshimiExp2015,ChangExp2016,GrauerExp2017,XueExp2018,ZhangExp2020,QAHIexp2020,ZhaoExp2020,MogiExp2021}, a band gap on the surface states of the topological materials is opened by the time-reversal symmetry breaking, which is essential for the realization of the quantum anomalous Hall insulator (QAHI) phase~\cite{XueExp2013,Qi2008,Yu2010,Chu2011}.
Physically, how to realize the QAHI phase has attracted much attention in the past few decades~\cite{Qiao2010,Qi2006,Onoda2006,Nomura2011,Checkelsky2012}. One of the promising physical schemes is on the basis of the topological insulators doped with magnetic impurities~\cite{Hor2010,Chen2010,Wray2011,Wang2013,Wang2014,Law2017,Kim2018,Kawamura2018}, where the interplay of the magnetic exchange interaction and the spin-orbit coupling gives rise to the band inversion between the conduction and valence bands.
More importantly, the experimental realization of the QAHI phase has been reported in the thin films of Cr-doped (Bi,Sb)$_2$Te$_3$, which is an intrinsic magnetic topological insulator~\cite{XueExp2013}.

The quantum spin Hall insulator (QSHI) phase is another family member of the Hall effects~\cite{Kane2005,Bernevig2006,Haldane2006,Prodan2009,Xing2010,Xing2011,Chen2017}. To explain QSHI, we take the Kane and Mele model~\cite{Kane2005} for example. In this model, the spin-up and -down
electrons exhibit opposite Chern numbers due to the spin-orbit coupling, so that the total Chern number vanishes but the spin Chern number is nonzero. Particularly, the QSHI phase has been observed experimentally in HgTe~\cite{QSHIexp2006,QSHIexp2007}. Usually, the QSHI phase was considered to be protected by the time-reversal symmetry. However, it has been found that the QSHI phase with nonzero spin Chern numbers persists when the time-reversal symmetry is broken and this is called the time-reversal-symmetry-broken QSHI phase~\cite{Xing2011,Chen2017}.

Based on the Floquet theory that allows one to map a time-dependent problem into a stationary one, it has been shown that a periodic perturbation can induce topological phase transitions in a topological trivial insulator~\cite{Oka2009prb,Calve2015prb,Kitagawa2011prb,Lindner2011np,Rechtsman2013nature,Wang2018prl,Xu2021as,Zhu2014,lightexp2013,lightexp2016,Leon2013,Katan2013,Inoue2010,Kitagawa2010,Yan2016,Narayan2015,Saha2016,Chen20181,Chen20182,Pervishko2018,Kyriienko2019,Huang20prl,Hu20prl,Du21arxiv,Ning21arxiv}.
In particular, the quantum Hall effect induced by circularly polarized field in Dirac materials was predicted decade ago by Oka and Aoki~\cite{Oka2009prb}. Further, the theory given by Oka and Aoki was generalized for various topological insulators in the previous Refs.~\cite{Calve2015prb,Kitagawa2011prb,Lindner2011np,Rechtsman2013nature,Wang2018prl,Xu2021as,Zhu2014,lightexp2013,lightexp2016,Leon2013,Katan2013,Inoue2010,Kitagawa2010,Yan2016,Narayan2015,Saha2016,Chen20181,Chen20182,Pervishko2018,Kyriienko2019,Huang20prl,Hu20prl,Du21arxiv,Ning21arxiv}. Later the corresponding Floquet topological insulators have been observed by experiments~\cite{lightexp2013,lightexp2016}. Also, it has been shown that an intense high-frequency linearly polarized light can be used to manipulate the value of a gap in of a non-doped topological insulator thin film~\cite{Pervishko2018}. However, the linearly polarized light cannot include the contribution of $1/\omega$, so that they only estimated the terms proportional to $1/\omega^2$~\cite{Pervishko2018}, where $\omega$ is the frequency of the polarized light. Noteworthy, the circularly polarized light can include the contribution of $1/\omega$.
Therefore, the impact of high-frequency pumping with the circularly polarized light on the thin films of topological insulators needs to be studied comprehensively.

In this work, we investigate the topological phase transitions in the atomically thin flakes of an intrinsic magnetic topological insulator with high-frequency pumping. Based on the results, it is found that the intensity of the circularly polarized light can be used as a knob to drive a topological transition. Different from the situation in the absent of optical field, there exist four different phases: the normal insulator (NI) phase, the time-reversal-symmetry-broken QSHI phase, and two different QAHI phases with opposite Chern numbers $C=\pm1$, respectively. In particular, for a special given layer thickness, one can apparently detect the tendency towards light-induced band inversion upon increasing optical field intensity and passing through the QSHI phase region. This means that the energy gap of the surface states can be tuned by adjusting the intensity of the driving optical field in an experimentally accessible range.

The paper is organized as the following: In Sec.~\ref{2}, we
give the model Hamiltonian. In Sec.~\ref{3}, we introduce the Floquet theory for a time-periodic Hamiltonian. In Sec.~\ref{4}, we give the polarized light and Floquet Hamiltonian which is used in the following calculations.
Furthermore, we study the basis states at the $\Gamma$ point in Sec.~\ref{5}.
Moreover, we calculate the high-frequency pumping-induced topological properties of the effective Floquet Hamiltonian of the thin film in Sec.~\ref{6}. In addition, we give the light-induced topological phase diagram in Sec.~\ref{7}. Finally, we summarize in Sec.~\ref{8}.

\section{Model}\label{2}

We take the periodic boundary conditions in the $x-y$ plane such that $k_x$ and $k_y$ are good quantum numbers,
and denote the thickness of the thin film along $z$ direction as $L$.
In the basis $(|p1_{z}^{+},\uparrow\rangle , |p2_{z}^{-},\uparrow\rangle , |p1_{z}^{+},\downarrow\rangle , |p2_{z}^{-},\downarrow\rangle )$ which
are the hybridized states of Se $p_{z}$ orbital and Bi $p_{z}$ orbital, with even ($+$) and odd ($-$) parities, up ($\uparrow$) and down ($\downarrow$) spins, the low-energy three-dimensional Hamiltonian for Cr-doped Bi$_{2}$Se$_{3}$ is given by~\cite{Shan2010,Lu2010,Xing2010,Lu2013,Liu2019,Sun2020,Liu2010prb,Dabiri2021prb1,Dabiri2021prb2}
\begin{align}\label{eq:H}
H({\bf k}) = H_{0}({\bf k}) + H_{X}(z),
\end{align} where
\begin{align}\label{eq:H0}
H_{0}({\bf k}) &\!=\! \epsilon_{0}({\bf k})I_{4} \nonumber\\
&~\!+\!
\!\left(\!
  \begin{array}{cccc}
    \!M({\bf k})\sigma_{z} \!-\! iA_{1}\partial_{z}\sigma_{x} & A_{2}k_{-}\sigma_{x} \\
    A_{2}k_{+}\sigma_{x}  & M({\bf k})\sigma_{z} \!+\! iA_{1}\partial_{z}\sigma_{x}\!
  \end{array}
\!\right)\! \nonumber\\
&\!=\! \epsilon_{0}({\bf k})\tau_{0}\otimes\sigma_{0} \!+\! M({\bf k})\tau_{0}\otimes\sigma_{z} \!+\! \!A_{1}\!k_{z}\tau_{z}\otimes\sigma_{x} \nonumber\\
&~\!+\! A_{2}k_{x}\tau_{x}\otimes\sigma_{x} \!+\! A_{2}k_{y}\tau_{y}\otimes\sigma_{x}.
\end{align} Here $\sigma_{x,y,z}$ are the Pauli matrices for the orbital degree of freedom, $k_z=-i\partial_{z}$, $I_{4}$ is the $4\times 4$ identity matrix, $k_{\pm}=k_{x}\pm ik_{y}$, $\epsilon_{0}({\bf k})=C_{0}-D_{1}\partial_{z}^{2}+D_{2}(k_{x}^{2}+k_{y}^{2})$, $M({\bf k})=M_{0}+B_{1}\partial_{z}^{2}-B_{2}(k_{x}^{2}+k_{y}^{2})$, $C_0$, $D_i$, $M_0$, $B_i$, and $A_i$ are model parameters with $i=1,2$.
The parameters for Bi$_{2}$Se$_{3}$ are adopted as~\cite{ZhangExp2009,XueExp2013}: $C_{0}=-0.0068$ eV, $D_{1}=1.3$ eV\AA$^2$, $D_{2}=19.6$ eV\AA$^2$, $A_{1}=2.2$ eV\AA, $A_{2}=4.1$ eV\AA, $M_{0}=0.28$ eV, $B_{1}=10$ eV\AA$^2$, and $B_{2}=56.6$ eV\AA$^2$.
The exchange field reads~\cite{Lu2013,Liu2019}
\begin{align}\label{eq:HX}
H_{X}(z) = m_{0}\tau_{z}\otimes\sigma_{0},
\end{align}
where $m_{0}$ is the magnitude of the bulk magnetic moment~\cite{XueExp2013}, $\tau_{z}$ is the $z$ Pauli
matrix for the spin degree of freedom, $\sigma_{0}$ is a $2\times2$ unit matrix, and the magnetization energy along the $z$ direction is given by $m_{0}$, i.e.,
$m_{0}$ is the exchange field from the magnetic dopants.

\section{Floquet formula}\label{3}

The Floquet theory can be applied to a time-periodic Hamiltonian $H(t) = H(t + T )$ with the period
$T = 2\pi/\omega$ and the frequency $\omega$ of the light. By employing the Floquet theory, the wave function of the time-periodic
Schr\"odinger equation $i\partial_{t}\Psi(t) = H(t)\Psi(t)$, has the form $\Psi(t)=\sum_{m}\psi_{m}e^{-i(\epsilon/\hbar + m\omega)t}$,
where $\epsilon$ is the quasienergy and $m$ is an integer. With a Fourier series expansion, we find that $\sum_{m}H_{n,m}\psi_{m} = \epsilon \psi_{n}$, where
\begin{align}\label{eq:HFnm}
H_{n,m} = n\hbar\omega\delta_{n,m} + \frac{1}{T} \int_{0}^{T}  H(t) e^{i(n-m)\omega t}dt,
\end{align}
which is a block Hamiltonian of the Floquet state, $n$ and $m$ are integers. If $\Psi(t)$ is an eigenvector with the quasienergy $\epsilon$, $e^{in\omega t}\Psi(t)$ is also an eigenvector of the system with the quasienergy $\epsilon + n\hbar\omega$.

From Eq.~(\ref{eq:HFnm}), one can have
\begin{align}\label{eq:Hnm}
H_{n,m}=
\left(
  \begin{array}{ccccc}
    \cdots & \cdots & \cdots & \cdots & \cdots  \\
    \cdots & H_{-1,-1}  & H_{-1,0}  & H_{-1,1}  & \cdots \\
    \cdots & H_{0,-1}  & H_{0,0} & H_{0,1} & \cdots  \\
    \cdots & H_{1,-1}  & H_{1,0}  & H_{1,1} & \cdots\\
    \cdots & \cdots  & \cdots  & \cdots & \cdots
  \end{array}
\right),
\end{align} where $H_{n,m} =H_{-m,-n}$ with $n\neq m$.

\section{Polarized light and Floquet Hamiltonian}\label{4}

The time-dependent Hamiltonian can be experimentally introduced by normally illuminating with elliptically polarized light described by a time-varying gauge field (or the vector potential) ${\bf A}(t) = A( \sin(\omega t),  \sin(\omega t + \varphi) )$~\cite{Saha2016,Zhu2014}, which gives the optical field as ${\bf E}(t) = \partial{\bf A}(t)/\partial t = E_{0}( \cos(\omega t),  \cos(\omega t + \varphi) )$, where $E_{0}=A/\omega$ is the amplitude of the optical field, $\omega$ is the frequency of the optical field, and the phase $\varphi$ controls the polarization: when $\varphi = 0$ or $\pi$, the optical field is linearly polarized; when $\varphi=\pm\pi/2$, the optical field is circularly polarized; for example, ${\bf A}(t) = A( \sin(\omega t),  \cos(\omega t) )$ with $\varphi=\pi/2$; when $\varphi$ takes other values, the optical field is elliptically polarized. 
From the point of view of the source~\cite{wikipedia_Circular}, with $\varphi=\pi/2$, the corresponding electric field vector is ${\bf E}(t) = E_{0}(\cos(\omega t), \cos(\omega t + \pi/2)) = E_{0}(\cos(\omega t), -\sin(\omega t))$, which is a left-handed circularly polarized wave and corresponds to ${\bf A}(t) = A( \sin(\omega t),  \cos(\omega t) )$. For $\varphi=-\pi/2$, the corresponding electric field vector is ${\bf E}(t) = E_{0}(\cos(\omega t), \cos(\omega t - \pi/2)) = E_{0}(\cos(\omega t), \sin(\omega t))$, which is a right-handed circularly polarized wave and corresponds to ${\bf A}(t) = A( \sin(\omega t),  -\cos(\omega t) )$. The vector potentials for right- and left-handed circularly polarized lights are different from each other and different vector potential will give different time-dependent Hamiltonian (\ref{eq:Ht}). In the following calculations, we focus on the left-handed circularly polarized light with $\varphi=\pi/2$. 
As a high-frequency laser light, we choose the photon energy, for example, to be $\hbar\omega\approx0.15$ eV ($\omega \sim 2.2789\times10^{2}$ THz), which is close to the typical values in recent optical pump-probe experiments. For example, the typical amplitude of light $A_{0}=eA/\hbar=eE_{0}/(\hbar\omega)$ is about 0.03 \AA$^{-1}$,
and the corresponding electric field strength $E_{0}=\hbar\omega A_{0}/e$ is $4.5\times10^{7}$ V/m, which is within the  experimental accessibility~\cite{lightexp2013,lightexp2016}.

By use of the Peierls substitution, the time-dependent Hamiltonian is obtained as
\begin{align}\label{eq:Ht}
H(t)= H\left({\bf k}_{\perp} - \frac{e}{\hbar}{\bf A}(t), -i\partial_{z}\right) ,
\end{align} where ${\bf k}_{\perp} = (k_x, k_y)$.
Making use of the Floquet theory~\cite{Zhu2014,Pervishko2018,Chen20181,Chen20182} in the high-frequency limit, the periodically driven system can be described by a static effective Hamiltonian as~\cite{high-frequency1982,high-frequency1988,high-frequency20031,high-frequency20032,high-frequency2014,high-frequency20151,high-frequency20152}
\begin{align}\label{eq:HF0}
H^{(F)} =  H_{0,0} + \frac{[H_{0,-1}, H_{0,1}]}{\hbar\omega} ,
\end{align} where
\begin{align}\label{eq:H00}
H_{0,0}
= H({\bf k}) + D_{2}A_{0}^{2} - B_{2}A_{0}^{2}I_{2}\otimes\sigma_{z},
\end{align}
\begin{align}\label{eq:H0-1}
H_{0,-1}
&\!=\! -iD_{2}A_{0}(k_{x} \!+\! e^{-i\varphi}k_{y}) \nonumber\\
&\!+\!
\!\left(\!
  \begin{array}{cccc}
    iB_{2}A_{0}(k_{x} \!+\! e^{-i\varphi}k_{y})\sigma_{z} & \!-\!\frac{i}{2}A_{2}A_{0}(1 \!-\! ie^{-i\varphi})\sigma_{x}\!  \\
    \!-\!\frac{i}{2}A_{2}A_{0}(1 \!+\! ie^{-i\varphi})\sigma_{x} & iB_{2}A_{0}(k_{x} \!+\! e^{-i\varphi}k_{y})\sigma_{z}\!
  \end{array}
\!\right)\!,
\end{align}
\begin{align}\label{eq:H01}
H_{0,1}
&\!=\! iD_{2}A_{0}(k_{x} + e^{i\varphi}k_{y}) \nonumber\\
&\!+\!
\!\left(\!
  \begin{array}{cccc}
    \!-\!iB_{2}A_{0}(k_{x} \!+\! e^{i\varphi}k_{y})\sigma_{z} & \frac{i}{2}A_{2}A_{0}(1 \!-\! ie^{i\varphi})\sigma_{x}  \\
    \frac{i}{2}A_{2}A_{0}(1 \!+\! ie^{i\varphi})\sigma_{x} & \!-\!iB_{2}A_{0}(k_{x} \!+\! e^{i\varphi}k_{y})\sigma_{z}\!
  \end{array}
\!\right)\!.
\end{align}
From Eq.~(\ref{eq:HF0}), the Floquet Hamiltonian is
\begin{align}\label{eq:HF}
H^{(F)} &\!=\!
\!\left(\!
  \begin{array}{cccc}
    \tilde{M}({\bf k})\sigma_{z} \!-\! iA_{1}\partial_{z}\sigma_{x} &    A_{2}k_{-} \sigma_{x} \\
    A_{2}k_{+} \sigma_{x}  & \tilde{M}({\bf k})\sigma_{z} \!+\! iA_{1}\partial_{z}\sigma_{x}\!
  \end{array}
\!\right)\! \nonumber\\
&\!+\! \frac{A_{0}^{2}A_{2}\sin\varphi}{\hbar\omega} \!\left(\!
  \begin{array}{cccc}
    \!-\!A_{2}I_{2} & \!-\! B_{2}k_{-}(\sigma_{+} \!-\! \sigma_{-})\!  \\
    \!B_{2}k_{+}(\sigma_{+} \!-\! \sigma_{-})  & A_{2}I_{2}
  \end{array}
\!\right)\! \nonumber\\
&\!+\! \tilde{\epsilon}_{0}({\bf k})I_{4} ,
\end{align} where $\tilde{\epsilon}_{0}({\bf k})=\epsilon_{0}({\bf k}) + A_{0}^{2}D_{2}$, $\tilde{M}({\bf k})=M({\bf k}) - A_{0}^{2}B_{2}$, $\sigma_{+}=\sigma_{x} + i\sigma_{y}$, and $\sigma_{-}=\sigma_{x} - i\sigma_{y}$.

To give a simple quantitative estimation of the validity of the theoretical formalism
developed here, we evaluate the maximum instantaneous energy of the time-dependent Hamiltonian (\ref{eq:Ht}) averaged over
a period of the field $\frac{1}{T}\int_{0}^{T}dt~\text{max}\left\{\big|\big|H(t)\big|\big|\right\}<\hbar\omega$.
Therefore, in the vicinity of the $\Gamma$ point, the field parameters have to meet the condition $A_{0}A_{2}/(\hbar\omega)<1$. Particularly, in the high-frequency regime for an external pumping $\hbar\omega\approx0.4$ eV ($\omega \sim 6.07707\times10^{2}$ THz), one can estimate $A_{0}\lesssim0.09$ \AA$^{-1}$.

Experimentally, from the spectroscopic ellipsometry measurements of the bulk Bi$_2$Se$_3$~\cite{McIver2012}, the penetration depth of the laser light is about $25$ nm~\cite{McIver2012}. 
At the same time, from the transient reflectivity/transmission measurements of the bulk Bi$_2$Se$_3$~\cite{Glinka2013}, the penetration depth of the laser light is about $21$ nm~\cite{Glinka2013}.
Both $21$ nm and $25$ nm are reasonably larger than the film thickness ($2\sim8$ nm) of the bulk Bi$_2$Se$_3$.
It is important that the penetration depth of the laser light should be reasonably larger than the film thickness, so that the vector potential ${\bf A}(t)$ of the laser light can properly change the momentum ${\bf k}_{\perp}$ as in the thin-film Hamiltonian~(\ref{eq:Ht}). Otherwise, the effect of the vector potential will not go into the bulk state of the thin film due to the skin effect.

\section{Basis states at the $\Gamma$ point}\label{5}

To establish an effective model for the surface states, we
first find the four solutions to the surface states of the model
in Eq.~(\ref{eq:H}) at the $\Gamma$ point ($k_x=k_y= 0$) as~\cite{Shan2010,Lu2010}
\begin{align}\label{eq:H02}
H_{0} =
\left(
  \begin{array}{cccc}
    h(A_{1}) & 0 \\
    0 & h(-A_{1})
  \end{array}
\right),
\end{align} where
\begin{align}
h(A_{1})
&\!=\! (C_{0}-D_{1}\partial_{z}^{2})I_{2}+(M_{0}+B_{1}\partial_{z}^{2})\sigma_{z} - iA_{1}\partial_{z}\sigma_{x} \nonumber\\
&\!=\!
\left(
  \begin{array}{cccc}
    C_{0}+M_{0}-D_{-}\partial_{z}^{2} & -iA_{1}\partial_{z} \\
   -i A_{1}\partial_{z} & C_{0}-M_{0}-D_{+}\partial_{z}^{2}
  \end{array}
\right),
\end{align} and $D_{\pm}=D_{1}\pm B_{1}$.

$H_{0}$ in Eq.~(\ref{eq:H02}) is block-diagonal and its solution can be found by solving each block separately, i.e.,
\begin{align}
&h(A_{1}) \Psi_{\uparrow}(z) = E \Psi_{\uparrow}(z),\label{eq:up} \\
&h(-A_{1}) \Psi_{\downarrow}(z) = E \Psi_{\downarrow}(z).
\end{align} Because the lower block is the ``time'' reversal of the upper block, the solutions satisfy $\Psi^{\downarrow}(z) = \Theta\Psi^{\uparrow}(z)$, where
$\Theta=-i\sigma_{y}\mathcal{K}$ is the time-reversal operator and $\mathcal{K}$ is the complex conjugation operation. Equivalently, we can replace $A_1$ by $-A_1$ in all the results for the upper block, to obtain those for the lower block. Therefore, we only need to solve $h(A_1)$.

The solution of the block-diagonal $H_{0}$ can be found by putting a two-component trial solution into the eigenequation (\ref{eq:up}) of the upper block with $\Psi_{\uparrow}(z) = \Psi^{\uparrow}_{\lambda}(z) e^{\lambda z}$, where $\lambda$ is the trial coefficients defining the behavior of the wave functions and $E$ is the trial eigenenergy.
Therefore, one can have
\begin{align}
E_{\pm} &= C_{0} - D_{1}\lambda^{2} \nonumber\\
&~~\pm \sqrt{(M_{0} + B_{1}\lambda^{2} - A_{1}\lambda ) (M_{0} + B_{1}\lambda^{2} + A_{1}\lambda )},
\end{align} and
\begin{align}\label{eq:eigenvector1}
\Psi^{\uparrow}_{\lambda} = \left(
  \begin{array}{cc}
     D_{+}\lambda^{2} - L_{-} + E_{\pm}  \\
    -i A_{1}\lambda
  \end{array}
\right).
\end{align}

Note that the trial coefficients may have multiple solutions, the final solution should be a linear superposition of these solutions with the superposition coefficients determined by boundary conditions. Then the problem becomes a straightforward calculation of the Schr\"odinger equation or the secular equation $\text{det}|h(A_{1}) - E|=0$ which gives four solutions of $\lambda_{\alpha}(E)$, denoted as $\beta\lambda_{\alpha}(E)$, with $\alpha\in \{1, 2\}$, $\beta\in \{+, -\}$ and
$\lambda_{\alpha}$ define the behavior of the wave functions along $z$ axis and are functions of the energy $E$ as
\begin{align}
\!\lambda_{\alpha}(E) \!=\! \sqrt{\frac{\!-\! F \!+\! (-1)^{\alpha-1}\sqrt{R}}{2D_{+}D_{-}}},
\end{align}
where we have defined
\begin{align}
F&=A_{1}^{2} + 2D_{1}(E - C_{0}) - 2B_{1}M_{0}\nonumber\\
&=A_{1}^{2} + D_{+}(E - L_{+}) + D_{-}(E - L_{-}), \\
R&=F^{2} - 4(D_{1}^{2} -B_{1}^{2})[(E-C_{0})^{2} - M_{0}^{2}]\nonumber\\
&=F^{2} - 4D_{+} D_{-}(E - L_{+}) (E - L_{-}),
\end{align} $D_{\pm}=D_{1}\pm B_{1}$, and $L_{\pm}=C_{0}\pm M_{0}$.
With Eq.~(\ref{eq:eigenvector1}), the general solution is a linear combination of the four linearly independent two-component vectors
\begin{align}\label{eq:wavefunctionup1}
\Psi_{\uparrow}(z) &= \sum_{\alpha=1,2} \sum_{\beta=+,-} C_{\alpha\beta} \Psi^{\uparrow}_{\alpha\beta} e^{\beta\lambda_{\alpha}z} \nonumber\\
&= \sum_{\alpha=1,2} \sum_{\beta=+,-} C_{\alpha\beta} \left(
  \begin{array}{cc}
     D_{+}\lambda_{\alpha}^{2} - L_{-} + E_{\pm}  \\
    -i A_{1}(\beta\lambda_{\alpha} )
  \end{array}
\right) e^{\beta\lambda_{\alpha}z},
\end{align} where the superposition coefficients $C_{\alpha\beta}$ are determined by boundary conditions.

\section{Effective model of the thin film}\label{6}

In this section, we derive the effective low-energy continuous model for the thin film of the three-dimensional topological insulators.

\subsection{Finite-thickness boundary conditions}\label{sec:Finite-thickness}\label{6.1}

Now, we turn to discuss the gap in a thin film with both top and bottom open surfaces.
When the thickness $L$ of the film is comparable with the characteristic length $1/\lambda$ of the
surface states, there is a coupling between the states on opposite surfaces. One has to consider
the boundary conditions at both surfaces simultaneously. Without loss of generality, we will
consider that the top surface is located at $z = L/2$ and the bottom surface at $-L/2$. The
boundary conditions are given as
\begin{align}\label{eq:finite-thickness}
\Psi_{\uparrow}\left(z=\pm\frac{L}{2} \right) = 0,
\end{align} where $L=N_{L}d$ is the thickness of the film with the number $N_{L}$ of the layers and the thickness $d$ of a layer, and $-L/2\leqslant z \leqslant L/2$.

With the boundary conditions in Eq.~(\ref{eq:finite-thickness}), the trial wave function can be given by
\begin{align}\label{eq:finite-wavefunction}
\Psi_{\uparrow}(z) =  \left(
  \begin{array}{cc}
    \psi_{1}(z)    \\
    \psi_{2}(z)
  \end{array}
\right),
\end{align} where
\begin{align}
\psi_{1}(z) &=  c_{+}f_{+}(z) + c_{-}f_{-}(z), \\
\psi_{2}(z) &=  d_{+}f_{+}(z) + d_{-}f_{-}(z),
\end{align}
\begin{align}
f_{+}(z) &= \frac{\cosh(\lambda_{1}z)}{\cosh(\lambda_{1}L/2)} - \frac{\cosh(\lambda_{2}z)}{\cosh(\lambda_{2}L/2)}, \\
f_{-}(z) &=  \frac{\sinh(\lambda_{1}z)}{\sinh(\lambda_{1}L/2)} - \frac{\sinh(\lambda_{2}z)}{\sinh(\lambda_{2}L/2)}.
\end{align}
Substituting Eq.~(\ref{eq:finite-wavefunction}) into (\ref{eq:up}): $h(A_{1})\Psi_{\uparrow}(z) \!=\! E \Psi_{\uparrow}(z)$, we have
\begin{align}\label{eq:finite-eigenequation}
\!\left(\!
  \begin{array}{cccc}
    \!C_{0}\!+\!M_{0}\!-\!E\!-\!D_{-}\partial_{z}^{2}\! & -iA_{1}\partial_{z} \\
   -i A_{1}\partial_{z} & \!C_{0}\!-\!M_{0}\!-\!E\!-\!D_{+}\partial_{z}^{2}\!
  \end{array}
\!\right)\! \!\left(\!
  \begin{array}{cc}
    \!\psi_{1}(z)\!    \\
    \!\psi_{2}(z)\!
  \end{array}
\!\right)\!=\!0.
\end{align}
With Eq.~(\ref{eq:finite-eigenequation}), we have
\begin{align}
\frac{c_{+}}{d_{-}} &= \frac{C_{0}-M_{0}-E-D_{+}\lambda_{1}^{2}}{ i A_{1}\lambda_{1}} \frac{\cosh(\lambda_{1}L/2)}{\sinh(\lambda_{1}L/2)}, \\
\frac{c_{+}}{d_{-}} &= \frac{C_{0}-M_{0}-E-D_{+}\lambda_{2}^{2}}{ i A_{1}\lambda_{2}} \frac{\cosh(\lambda_{2}L/2)}{\sinh(\lambda_{2}L/2)}, \\
\frac{c_{-}}{d_{+}} &= \frac{C_{0}-M_{0}-E-D_{+}\lambda_{1}^{2}}{ i A_{1}\lambda_{1}} \frac{\sinh(\lambda_{1}L/2)}{\cosh(\lambda_{1}L/2)}, \\
\frac{c_{-}}{d_{+}} &= \frac{C_{0}-M_{0}-E-D_{+}\lambda_{2}^{2}}{ i A_{1}\lambda_{2}} \frac{\sinh(\lambda_{2}L/2)}{\cosh(\lambda_{2}L/2)}.
\end{align}
Furthermore, the secular equation of the nontrivial solution to the superposition coefficients $C_{\alpha\beta}$ leads to the transcendental equations
\begin{align}
&\frac{[C_{0} \!-\! M_{0} \!-\! E_{+}^{0} \!-\! D_{+}\lambda_{1}^{2}(E_{+}^{0}) ]\lambda_{2}(E_{+}^{0})}{[C_{0} \!-\! M_{0} \!-\! E_{+}^{0} \!-\! D_{+}\lambda_{2}^{2}(E_{+}^{0}) ]\lambda_{1}(E_{+}^{0})} \!=\! \frac{\tanh(\lambda_{2}(E_{+}^{0})L/2)}{\tanh(\lambda_{1}(E_{+}^{0})L/2)}, \label{eq: transcendentalequation21} \\
&\frac{[C_{0} \!-\! M_{0} \!-\! E_{-}^{0} \!-\! D_{+}\lambda_{2}^{2}(E_{-}^{0}) ]\lambda_{1}(E_{-}^{0})}{[C_{0} \!-\! M_{0} \!-\! E_{-}^{0} \!-\! D_{+}\lambda_{1}^{2}(E_{-}^{0}) ]\lambda_{2}(E_{-}^{0})} \!=\! \frac{\tanh(\lambda_{2}(E_{-}^{0})L/2)}{\tanh(\lambda_{1}(E_{-}^{0})L/2)}. \label{eq: transcendentalequation22}
\end{align}

\begin{figure}[htpb]
\centering
\includegraphics[width=0.47\textwidth]{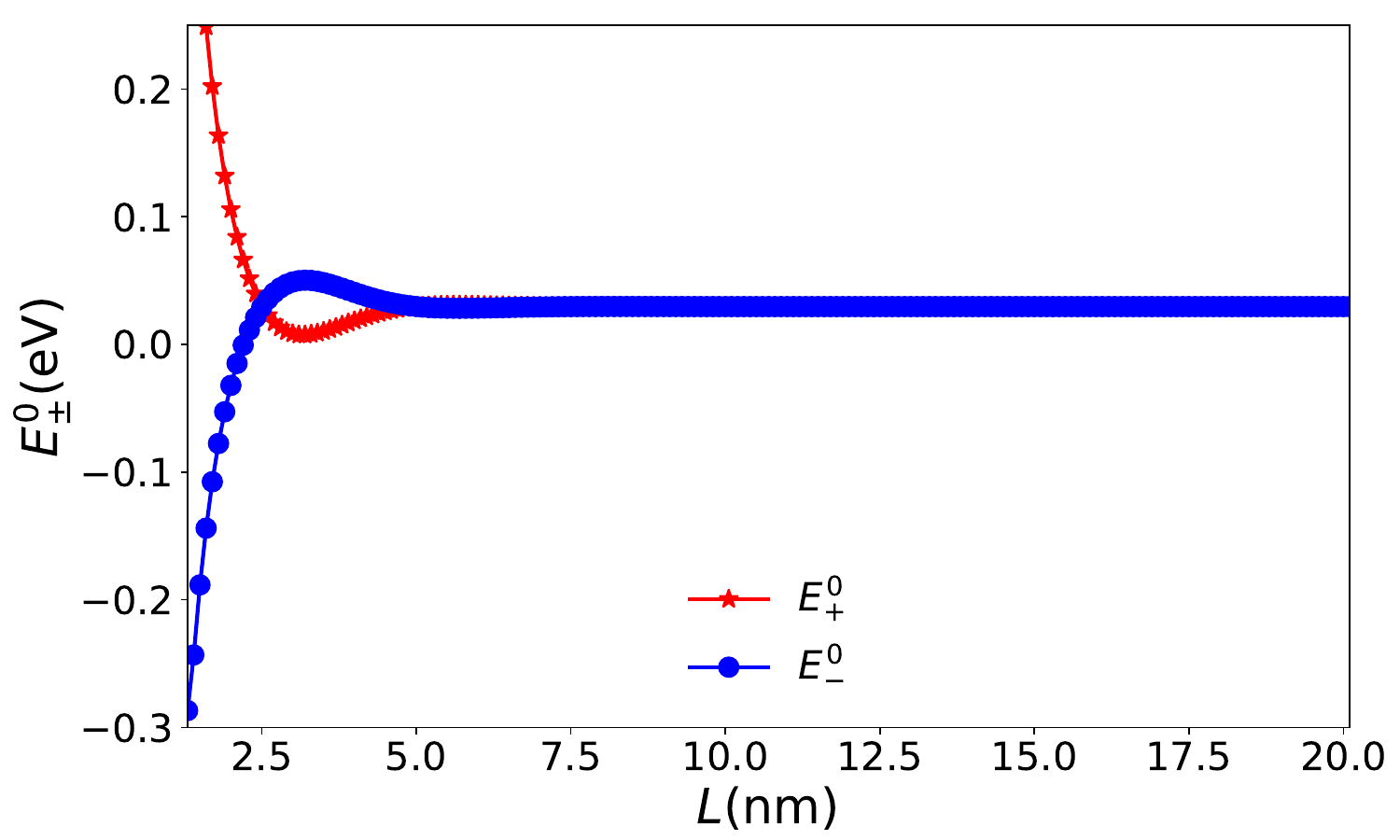}
\caption{The eigen energies $E_{+}^{0}$ and $E_{-}^{0}$ as a function of the finite thickness $L$ for Bi$_{2}$Se$_{3}$. The red solid star points are for $E_{+}^{0}$ , and the blue solid circle points are for $E_{-}^{0}$. The numerical solutions are obtained by solving the transcendental equations (\ref{eq: transcendentalequation21}) and (\ref{eq: transcendentalequation22}).} \label{Fig:E0zf}
\end{figure}

The numerical solutions of the transcendental equations (\ref{eq: transcendentalequation21}) and (\ref{eq: transcendentalequation22}) are shown in Fig.~\ref{Fig:E0zf}.

In particular, for $L\rightarrow+\infty$ (infinitely thick case), we have the analytic solutions for Eqs.(\ref{eq: transcendentalequation21}) and (\ref{eq: transcendentalequation22}) as (the detail derivations can be found in the Appendix A)
\begin{align}\label{eq:E0zf}
E_{\pm}^{0}\xrightarrow{L\rightarrow+\infty} C_{0} + \frac{D_{1}M_{0} }{B_{1} } = 0.0296~({\rm eV}),
\end{align} which is consistent with the numerical results shown in Fig.~\ref{Fig:E0zf}. 

For the infinitely thin case ($L\rightarrow0$), there will be no surface state which corresponds to no boundary condition. Besides, there is no wave function in the bulk and surface of the thin film physically. Therefore, there is no analytic solutions in the infinitely thin case.

Therefore, the eigen wavefunctions for $E_{+}^{0}$ and $E_{-}^{0}$ are, respectively,
\begin{align}\label{eq:wavefunctionup-finite}
\varphi(A_{1}) \equiv \Psi_{\uparrow}^{+} &= \bar{C}_{+} \left(
  \begin{array}{cc}
    - D_{+}\eta_{1}^{+}f_{-}^{+}   \\
    i A_{1}f_{+}^{+}
  \end{array}
\right),  \\
\chi(A_{1}) \equiv \Psi_{\uparrow}^{-} &= \bar{C}_{-} \left(
  \begin{array}{cc}
    - D_{+}\eta_{2}^{-}f_{+}^{-}   \\
    i A_{1}f_{-}^{-}
  \end{array}
\right),
\end{align} where $\bar{C}_{\pm}$ is the normalization factor. The superscripts of $f_{\pm}^{\pm}$ and $\eta_{1,2}^{\pm}$ stand for $E_{\pm}^{0}$
and the subscripts of $f_{\pm}^{\pm}$ for parity, respectively. The expressions for $f_{\pm}^{\pm}$ and $\eta_{1,2}^{\pm}$ are given by
\begin{align}
f_{+}^{\pm}(z) &= \left[ \frac{\cosh(\lambda_{1}z)}{\cosh(\lambda_{1}L/2)} - \frac{\cosh(\lambda_{2}z)}{\cosh(\lambda_{2}L/2)} \right]_{E=E_{\pm}^{0}}, \\
f_{-}^{\pm}(z) &=  \left[ \frac{\sinh(\lambda_{1}z)}{\sinh(\lambda_{1}L/2)} - \frac{\sinh(\lambda_{2}z)}{\sinh(\lambda_{2}L/2)} \right]_{E=E_{\pm}^{0}} , \\
\eta_{1}^{\pm} &= \frac{\lambda_{1}^{2} - \lambda_{2}^{2}}{\lambda_{1}\coth(\lambda_{1}L/2) - \lambda_{2}\coth(\lambda_{2}L/2) } \bigg|_{E=E_{\pm}^{0}}, \\
\eta_{2}^{\pm} &= \frac{\lambda_{1}^{2} - \lambda_{2}^{2}}{\lambda_{1}\tanh(\lambda_{1}L/2) - \lambda_{2}\tanh(\lambda_{2}L/2) } \bigg|_{E=E_{\pm}^{0}} .
\end{align}

\subsection{Effective Floquet Hamiltonian of the thin film}\label{6.2}

\begin{figure}[htpb]
\centering
\includegraphics[width=0.47\textwidth]{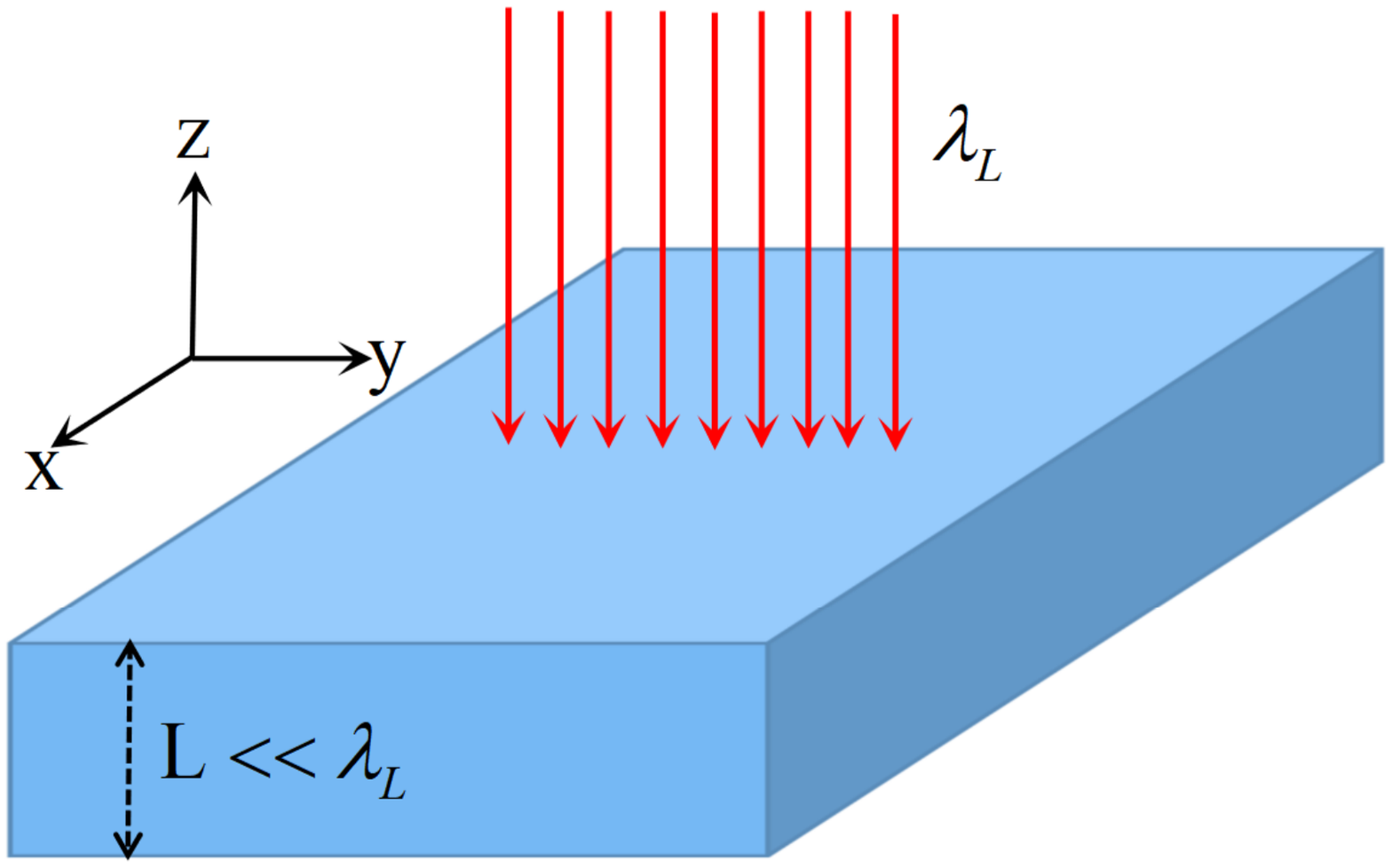}
\caption{A thin film of Bi$_{2}$Se$_{3}$ pumped by a circularly polarized optical field propagating to its surface. Here we suppose that $L\ll\lambda_{L}$ with the thickness $L$ of the thin film and the wavelength $\lambda_{L}$ of the driving optical field.} \label{Fig:Fig1}
\end{figure}

The energy spectra and wavefunctions of the lower block $h(A_{1})$ of $H_{0}$ can be obtained directly
by replacing $A_{1}$ by $- A_{1}$. Based on the above discussions, the four eigenstates of $H_{0}$ can be given by
\begin{align}
\Phi_{1}&=  \left(
  \begin{array}{cc}
    \varphi(A_{1})    \\
    0
  \end{array}
\right),~~~~~~~\Phi_{2} =  \left(
  \begin{array}{cc}
    \chi(A_{1})    \\
    0
  \end{array}
\right), \\
\Phi_{3} &=  \left(
  \begin{array}{cc}
    0    \\
    \varphi(-A_{1})
  \end{array}
\right),~~~~~\Phi_{4} =  \left(
  \begin{array}{cc}
    0   \\
    \chi(-A_{1})
  \end{array}
\right),
\end{align} with $\Phi_{1}\rightarrow\Phi_{3}$ and $\Phi_{2}\rightarrow\Phi_{4}$ under the time-reversal operation.
We should emphasize that these four solutions are for the surface states, and we use the four states as the basis states.

With the help of the four states, at the $\Gamma$ point, we can expand the Hamiltonian
Eq.~(\ref{eq:HF}) to obtain a new Floquet Hamiltonian of the thin film
\begin{align}\label{eq:HFfilm}
\!H_{\text{film}}^{(F)} &\!=\! \int_{\!-\!L/2}^{L/2}\!dz\!\{\Phi_{1}, \Phi_{4}, \Phi_{2}, \Phi_{3}\}^{\dagger} [\!H^{(F)}\!(\!{\bf k}\!)\!] \{\Phi_{1}, \Phi_{4}, \Phi_{2}, \Phi_{3}\} \nonumber\\
&\!=\! \left(
  \begin{array}{cccc}
    h_{f1}(k_{\perp}) & 0  \\
    0 & h_{f2}(k_{\perp})
  \end{array}
\right),
\end{align}
where
\begin{align}
\!h_{f1}(k_{\perp}) &\!=\! \tilde{E}_{1}^{0} \!-\! Dk_{\perp}^{2} \!+\! \!\left(\!
  \begin{array}{cccc}
    \frac{\tilde{\Delta}}{2} \!-\! Bk_{\perp}^{2} \!+\! \tilde{m} &  i\gamma_{1} k_{-}  \\
    - i\gamma_{1} k_{+} & - \frac{\tilde{\Delta}}{2} \!+\! Bk_{\perp}^{2} \!-\! \tilde{m}
  \end{array}
\!\right)\! \nonumber\\
&\!=\! (\tilde{E}_{1}^{0} \!-\! Dk_{\perp}^{2})\sigma_0 \!+\! \left(\frac{\tilde{\Delta}}{2} \!-\! Bk_{\perp}^{2} \!+\! \tilde{m} \right)\sigma_z  \nonumber\\
&~\!+\! i\gamma_{1} k_{-}\sigma_{+} \!-\! i\gamma_{1} k_{+}\sigma_{-}, \\
\!h_{f2}(k_{\perp}) &\!=\! \tilde{E}_{2}^{0} \!-\! Dk_{\perp}^{2} \!+\! \!\left(\!
  \begin{array}{cccc}
     -\frac{\tilde{\Delta}}{2} \!+\! Bk_{\perp}^{2} \!+\! \tilde{m}  &  i\gamma_{2} k_{-}  \\
     - i\gamma_{2} k_{+}  & \frac{\tilde{\Delta}}{2} \!-\! Bk_{\perp}^{2} \!-\! \tilde{m}
  \end{array}
\!\right)\! \nonumber\\
&\!=\! (\tilde{E}_{2}^{0} \!-\! Dk_{\perp}^{2})\sigma_0 \!+\! \left(-\frac{\tilde{\Delta}}{2} \!+\! Bk_{\perp}^{2} \!+\! \tilde{m} \right)\sigma_z  \nonumber\\
&~\!+\! i\gamma_{2} k_{-}\sigma_{+} \!-\! i\gamma_{2} k_{+}\sigma_{-},
\end{align} $\sigma_{\pm}=(\sigma_{x}\pm i\sigma_{y})/2$, $\tilde{E}_{1}^{0}\!=\!E^{0} \!-\! A_{0}^{2}D \!+\! \frac{m_{1} \!-\! m_{3}}{2}$, $\tilde{E}_{2}^{0}\!=\!E^{0} \!-\! A_{0}^{2}D \!-\! \frac{m_{1} \!-\! m_{3}}{2}$, $\frac{\tilde{\Delta}}{2}\!=\!\frac{\Delta}{2} \!-\! A_{0}^{2}B$, $\tilde{m}\!=\!\frac{m_{1} \!+\! m_{3}}{2} \!-\! A_{0}^{2}A_{2}^{2}\sin\varphi/(\hbar\omega)$, $E^{0}\!=\!(E_{+}^{0} \!+\! E_{-}^{0})/2$, $\Delta\!=\!E_{+}^{0} \!-\! E_{-}^{0}$, $B\!=\!(\tilde{B}_{1} \!-\! \tilde{B}_{2})/2$, $D\!=\!(\tilde{B}_{1} \!+\! \tilde{B}_{2})/2 \!-\! D_{2}$, $$\tilde{B}_{1} \!=\! B_{2}|\bar{C}_{+}|^{2} \int_{-L/2}^{L/2}dz \left( D_{+}^{2}|\eta_{1}^{+}f_{-}^{+}|^{2} \!-\! A_{1}^{2}|f_{+}^{+}|^{2}  \right),$$ $$\tilde{B}_{2} \!=\! B_{2}|\bar{C}_{-}|^{2} \int_{-L/2}^{L/2}dz \left( D_{+}^{2}|\eta_{2}^{-}f_{+}^{-}|^{2} \!-\! A_{1}^{2}|f_{-}^{-}|^{2}\right),$$ $\gamma_{1}\!=\!\gamma \!-\! \gamma_{f}A_{0}^{2}\sin\varphi/(\hbar\omega)$, $\gamma_{2}\!=\!\gamma \!+\! \gamma_{f}A_{0}^{2}\sin\varphi/(\hbar\omega)$, $\gamma \!=\! A_{1}A_{2}\bar{C}_{+}^{*}\bar{C}_{-}D_{+} \int_{-L/2}^{L/2}dz \left( \eta_{1}^{+}f_{-}^{+}f_{-}^{-} \!+\! \eta_{2}^{-}f_{+}^{+}f_{+}^{-} \right)$, $\gamma_{f} = 2A_{1}A_{2}B_{2}\bar{C}_{+}^{*}\bar{C}_{-}D_{+} \int_{-L/2}^{L/2}dz \left( \eta_{1}^{+}f_{-}^{+}f_{-}^{-}  -  \eta_{2}^{-}f_{+}^{-}f_{+}^{+}  \right)$, $m_{1} \!=\! m_{0}|\bar{C}_{+}|^{2} \int_{-L/2}^{L/2}dz~\left( D_{+}^{2}|\eta_{1}^{+}f_{-}^{+}|^{2}  + A_{1}^{2}|f_{+}^{+}|^{2}  \right) \!=\! m_{0}$, and $m_{3} \!=\! m_{0}|\bar{C}_{-}|^{2} \int_{-L/2}^{L/2}dz~\left( D_{+}^{2}|\eta_{2}^{-}f_{+}^{-}|^{2} \!+\! A_{1}^{2}|f_{-}^{-}|^{2} \right) \!=\! m_{0}$.

From the time-dependent Hamiltonian (\ref{eq:Ht}), it is found that the circularly polarized light is acting as an effective ``magnetic field''. Furthermore, from the Floquet thin-film Hamiltonian (\ref{eq:HFfilm}), we find that the ``effective field'' is causing kind of a Zeeman effect with the term $\tilde{m}\tau_{0}\otimes\sigma_{z}$ in (\ref{eq:HFfilm}) and an effective Zeeman field $\tilde{m}\!=\!m_{0}\!-\! A_{0}^{2}A_{2}^{2}\sin\varphi/(\hbar\omega)$.

The dispersions are
\begin{align}
E_{f1\pm}&\!=\!\tilde{E}_{1}^{0} \!-\! Dk_{\perp}^{2} \!\pm\! \sqrt{ \left(\frac{\tilde{\Delta}}{2} \!-\! Bk_{\perp}^{2} \!+\! \tilde{m}\right)^{2} \!+\! \gamma_{1}^{2}k_{\perp}^{2} }, \\
E_{f2\pm}&\!=\!\tilde{E}_{2}^{0} \!-\! Dk_{\perp}^{2} \!\pm\! \sqrt{ \left(\frac{\tilde{\Delta}}{2} \!-\! Bk_{\perp}^{2} \!-\! \tilde{m}\right)^{2} \!+\! \gamma_{2}^{2}k_{\perp}^{2} }.
\end{align}
At the $\Gamma$ point with $k_{\perp}=0$, the surface gap $\Delta_{\text{sur}}$ is defined as the minimum energy gap between the conduction and valence bands, i.e.,
\begin{align}\label{eq:sur}
\Delta_{\text{sur}}\!=\!\text{Min}\!\left[\!2\bigg|\frac{\tilde{\Delta}}{2} \!+\! \tilde{m}\bigg|,2\bigg|\frac{\tilde{\Delta}}{2} \!-\! \tilde{m}\bigg|,\left(\!\bigg|\frac{\tilde{\Delta}}{2} \!+\! \tilde{m}\bigg|\!+\!\bigg|\frac{\tilde{\Delta}}{2} \!-\! \tilde{m}\bigg|\!\right) \!\right]\!,
\end{align} where we use $m_{1}=m_{3}=m_{0}$.
Further, the wavefunctions of the two valence bands $E_{f1-}$ and $E_{f2-}$ are found to be
\begin{align}
\psi_{f1-} \!=\! G_{f1}\!\left(\!
  \begin{array}{cccc}
    b_{f1}(k_{\perp})   \\
    -i\gamma_{1} k_{+}  \\
    0   \\
    0
  \end{array}
\!\right),~
\psi_{f2-} \!=\! G_{f2}\!\left(\!
  \begin{array}{cccc}
    0  \\
    0  \\
    b_{f2}(k_{\perp})  \\
    i\gamma_{2} k_{+}
  \end{array}
\!\right)\! ,
\end{align} where
\begin{align}
\!b_{f1}(\!k_{\perp}\!)\! &\!=\! \frac{\tilde{\Delta}}{2} \!-\! Bk_{\perp}^{2} \!+\! \tilde{m} \!-\! \sqrt{ \left(\!\frac{\tilde{\Delta}}{2} \!-\! Bk_{\perp}^{2} \!+\! \tilde{m}\!\right)^{2} \!+\! \gamma_{1}^{2}k_{\perp}^{2} }, \\
\!b_{f2}(\!k_{\perp}\!)\! &\!=\! \frac{\tilde{\Delta}}{2} \!-\! Bk_{\perp}^{2} \!-\! \tilde{m} \!+\! \sqrt{ \left(\!\frac{\tilde{\Delta}}{2} \!-\! Bk_{\perp}^{2} \!-\! \tilde{m}\!\right)^{2} \!+\! \gamma_{2}^{2}k_{\perp}^{2} }, \\
\!G_{f1} &\!=\! \frac{1}{\sqrt{ b_{f1}^{2}(k_{\perp}) \!+\! \gamma_{1}^{2}k_{\perp}^{2} }},~
\!G_{f2} \!=\! \frac{1}{\sqrt{ b_{f2}^{2}(k_{\perp}) \!+\! \gamma_{2}^{2}k_{\perp}^{2} }}.
\end{align}

\subsection{Chern number of a thin film with high-frequency pumping}\label{7.3}

In principle, we can find the Hall conductance for each $h_{f1/f2}(k_{\perp})$.
Note that $h_{f1/f2}(k_{\perp})$ in Eq.~(\ref{eq:HFfilm}) can be explicitly written as
\begin{align}
h_{f1}(k_{\perp})&\!=\!h_{f+}(k_{\perp})
\!=\! \tilde{E}_{1}^{0} \!-\! Dk_{\perp}^{2} \!+\! \sum_{i=x,y,z}d_{i}\sigma_{i},\\
h_{f2}(k_{\perp})&\!=\!h_{f-}(k_{\perp})
\!=\! \tilde{E}_{2}^{0} \!-\! Dk_{\perp}^{2} \!+\! \sum_{i=x,y,z}d_{i}\sigma_{i},
\end{align} where the subscripts are $f1=f+$, $f2=f-$, $\sigma_{i}$ are the Pauli matrices and the $\textbf{d}(k_{\perp})$ vectors are
\begin{align}
d_{x} &\!=\! \!-\! \left(\!-\! \gamma \!\pm\! \frac{A_{0}^{2}}{\hbar\omega}\gamma_{f}\sin\varphi \right) k_{y},\label{eq:dfx}\\
d_{y} &\!=\! \left(\!-\! \gamma \pm \frac{A_{0}^{2}}{\hbar\omega}\gamma_{f}\sin\varphi \right) k_{x},\label{eq:dfy}\\
d_{z} &\!=\!\pm \left( \frac{\tilde{\Delta}}{2} \!-\! Bk_{\perp}^{2}\right) \!+\! \tilde{m}.\label{eq:dfz}
\end{align}

For the $2\times2$ Hamiltonian in terms of the $\textbf{d}(k_{\perp})$ vectors and Pauli matrices, the Kubo formula for the Hall conductance can be generally expressed as~\cite{Lu2010,Qi2006}
\begin{align}\label{eq:HallconductanceF1}
\sigma_{xy} = \frac{e^{2}}{2\hbar} \int \frac{d^{2}\textbf{k}}{(2\pi)^{2}} \frac{(f_{k,c} - f_{k,\nu})}{d^{3}} \epsilon_{\alpha\beta\gamma} \frac{\partial d_{\alpha}}{\partial k_{x}} \frac{\partial d_{\beta}}{\partial k_{y}} d_{\gamma},
\end{align} where we use $\textbf{k}=(k_x,k_y)$, $k=k_{\perp}$, $\epsilon_{\alpha\beta\gamma}$ is the Levi-Civita anti-symmetric tensor, $d=\sqrt{d_{x}^{2} + d_{y}^{2} + d_{z}^{2} }$ is the norm of $(d_{x}, d_{y}, d_{z})$, $\hbar=h/(2\pi)$ is the reduced Planck's constant, $-e$ is the electron charge, $f_{k,c/\nu} = 1/\{\exp[(\varepsilon_{c/\nu}(k) - \mu)/(k_{B}T)] + 1\}$ is the Fermi distribution function of the conduction ($c$) and valence ($\nu$) bands with the chemical potential $\mu$, the Boltzmann constant $k_{B}$, and the temperature $T$ .

At zero temperature and when the chemical potential $\mu$ lies between $(-|\Delta_{\text{sur}}|/2, |\Delta_{\text{sur}}|/2)$, the Fermi functions reduce to $f_{k,c}= 0$ and $f_{k,\nu}= 1$. By substituting Eqs.~(\ref{eq:dfx})-(\ref{eq:dfz}) into (\ref{eq:HallconductanceF1}), we arrive at
\begin{align}\label{eq:HallconductanceF2}
\!\sigma_{xy}^{(\pm)} &\!=\! \!-\! \left(\! \pm\frac{e^{2}}{2h} \!\right) \nonumber\\
&\!\times\!\int_{0}^{\infty}\! \frac{\left(\!-\! \gamma \!\pm\! \frac{A_{0}^{2}}{\hbar\omega}\gamma_{f}\sin\varphi \!\right)^{2}\!\left(\! \frac{\tilde{\Delta}}{2} \!+\! Bk^{2} \pm \tilde{m} \!\right)\! kdk}{\left[\! \left(\! \!-\! \gamma \pm \frac{A_{0}^{2}}{\hbar\omega}\gamma_{f}\sin\varphi \!\right)^{2}\!k^{2} \!+\! \left(\! \frac{\tilde{\Delta}}{2} \!-\! Bk^{2} \!\pm\! \tilde{m} \!\right)^{2} \!\right]^{3/2}} ,
\end{align} where $+$ ($-$) corresponds to pseudo-spin-$\uparrow$ (pseudo-spin-$\downarrow$) or $h_{f+}(k_{\perp})$ ($h_{f-}(k_{\perp})$).
By defining
\begin{align}
\!\cos\theta \!=\! \frac{\left( \frac{\tilde{\Delta}}{2} \!-\! Bk^{2} \!\pm\! \tilde{m} \right)}{\!\left[\! \left( \!-\! \gamma \!\pm\! \frac{A_{0}^{2}}{\hbar\omega}\gamma_{f}\sin\varphi \right)^{2}k^{2} \!+\! \left( \frac{\tilde{\Delta}}{2} \!-\! Bk^{2} \!\pm\! \tilde{m} \!\right)^{2}\!\right]^{1/2}\!},
\end{align}
Eq.~(\ref{eq:HallconductanceF2}) can be transformed into
\begin{align}\label{eq:HallconductanceF3}
\sigma_{xy}^{(\pm)} = \pm\frac{e^{2}}{2h} \int_{0}^{\infty} d k^{2} \frac{\partial \cos\theta}{\partial k^{2}} .
\end{align}

The value of $\cos\theta$ at $k= 0$ and $k\rightarrow \infty$ only depends on the signs of $\frac{\tilde{\Delta}}{2}\pm\tilde{m}$ and $B$, respectively. As a result, in the insulating regime $-|\Delta_{\text{sur}}|/2 \leqslant \mu \leqslant |\Delta_{\text{sur}}|/2$, we find that the anomalous Hall conductance for each hyperbola has the form
\begin{align}\label{eq:anomalousHallConductance}
\sigma_{xy}^{(\pm)} =- \left( \pm\frac{e^{2}}{2h} \right) [\text{sgn}(\Delta_{\pm}) + \text{sgn}(B)],
\end{align} where $\Delta_{\pm}=\Delta_{c} \pm \tilde{m}$,
$\Delta_{c} = \frac{\tilde{\Delta}}{2} = \frac{\Delta}{2} - A_{0}^{2}B$, and
$\tilde{m} = \frac{m_{1}+m_{3}}{2} - A_{0}^{2}A_{2}^{2}\sin\varphi/(\hbar\omega)$.
In particular, the information of the interplay between the circularly polarized light and the bulk magnetic moment can be extracted from the analytical expression for the anomalous Hall conductance Eq.~(\ref{eq:anomalousHallConductance}). The competition between the polarized light and the bulk magnetic moment can change the first sign function [$\text{sgn}(\Delta_{\pm})$] in Eq.~(\ref{eq:anomalousHallConductance}). This interplay has significant impact on the anomalous Hall conductance or the topological phase diagram of Cr-doped Bi$_{2}$Se$_{3}$ thin film as shown in Fig.~\ref{Fig:PhaseDiagram3D}(b).

Therefore, the spin Chern numbers of the valence bands are given by
\begin{align}
C_{\pm} = - \left( \pm\frac{1}{2} \right) [\text{sgn}(\Delta_{\pm}) + \text{sgn}(B)].
\end{align}
Further, the total Chern number is given by
\begin{align}
C=C_{+}+C_{-}.
\end{align}

Since a net magnetic moment should also break time-reversal symmetry, it is important that the excitation decreases the magnetic moment as an effective magnetic moment $\tilde{m}\!=\!m_{0}\!-\! A_{0}^{2}A_{2}^{2}\sin\varphi/(\hbar\omega)$, where we choose $\varphi=\pi/2$. Therefore, the effective magnetic moment decreases with the increasing of the light intensity $A_{0}$.

\section{Phase diagram}\label{7}

\begin{figure}[htpb]
\centering
\includegraphics[width=0.48\textwidth]{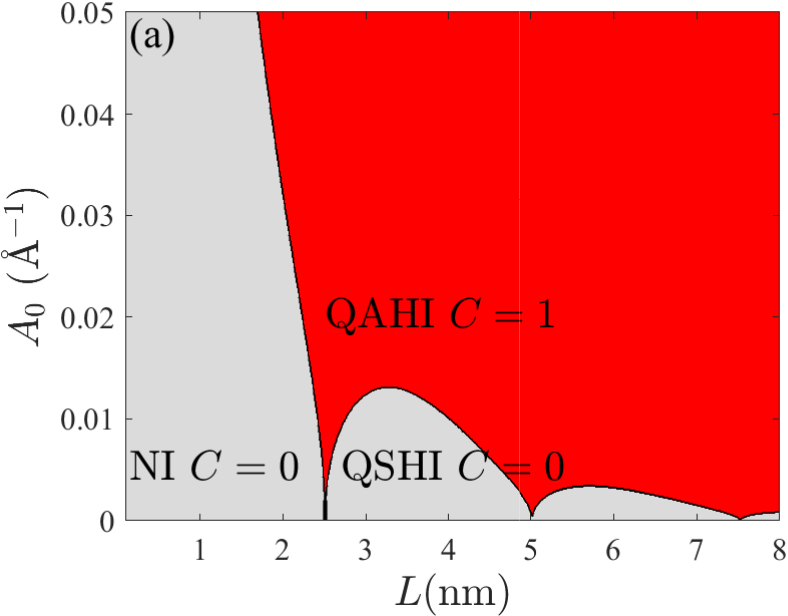} \\
\includegraphics[width=0.48\textwidth]{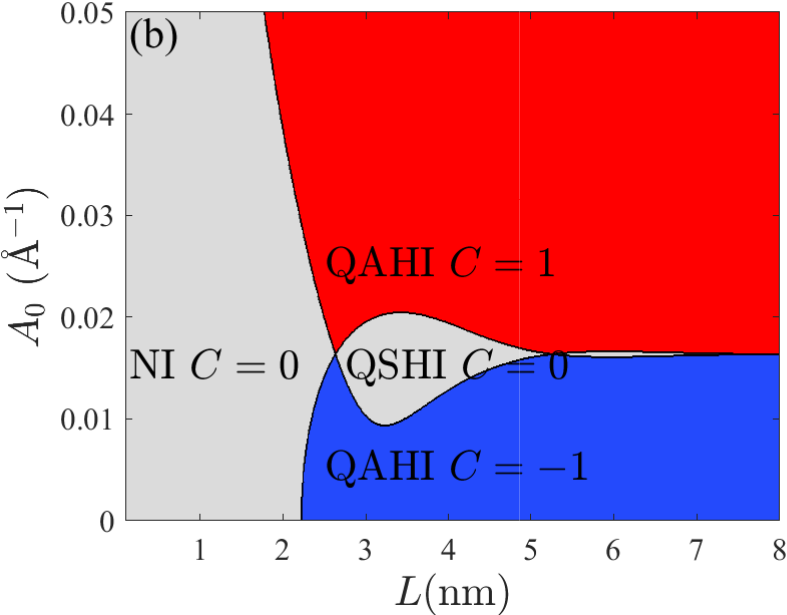}
\caption{The total Chern number as functions of the thin-film thickness and the light intensity.
(a) Topological phase diagram for the non-doped Bi$_{2}$Se$_{3}$ with $m_{0}=0$.
(b) Topological phase diagram for the Cr-doped Bi$_{2}$Se$_{3}$ with $m_{0}=30$ meV.
The other parameters are given as $\varphi=\pi/2$ and $\hbar\omega\approx150$ meV ($\omega \sim 2.2789\times10^{2}$ THz).} \label{Fig:PhaseDiagram3D}
\end{figure}

As shown in Fig.~\ref{Fig:PhaseDiagram3D}, we calculate the total Chern number as functions of the thin-film thickness and the light intensity for Bi$_{2}$Se$_{3}$.
For the non-doped Bi$_{2}$Se$_{3}$ with $m_{0}=0$ as shown in Fig.~\ref{Fig:PhaseDiagram3D}(a), by modulating the strength of the polarized optical field, there are three different regions: the NI phase with the spin Chern numbers $C_{\pm}=0$, the QAHI phase with the spin Chern numbers $C_{+}=1$ and $C_{-}=0$, and the QSHI phase with the spin Chern numbers $C_{\pm}=\pm1$. In the absent of optical field with $A_{0}=0$, there are only two phases: NI phase and QSHI phase. The result reveals that the circularly polarized light breaks the time-reversal symmetry and induces the QAHI phase. Moreover, the light-induced time-reversal-symmetry-broken QSHI phase is different from the time-reversal-symmetry QSHI phase in the absent of optical field.

For the Cr-doped Bi$_{2}$Se$_{3}$ with $m_{0}=30$ meV, it is indicated from Fig.~\ref{Fig:PhaseDiagram3D}(b) that there are four different regions: the NI phase, the time-reversal-symmetry-broken QSHI phase, and two different QAHI phases with opposite nonzero Chern numbers. Here the interplay between the light and bulk magnetic moment can separate two different QAHI phases with opposite Chern numbers.
For a given layer thickness $L=3.2$ nm, one can apparently detect the tendency towards light-induced band inversion upon increasing optical field intensity and passing through the QSHI phase region marked by gray.
In particular, the gray regions between the fixed points are the QSHI phases with the Chern number $C=0$. With increasing the thickness $L$, the gray regions will vanish.

\begin{figure}[htpb]
\centering
\includegraphics[width=0.48\textwidth]{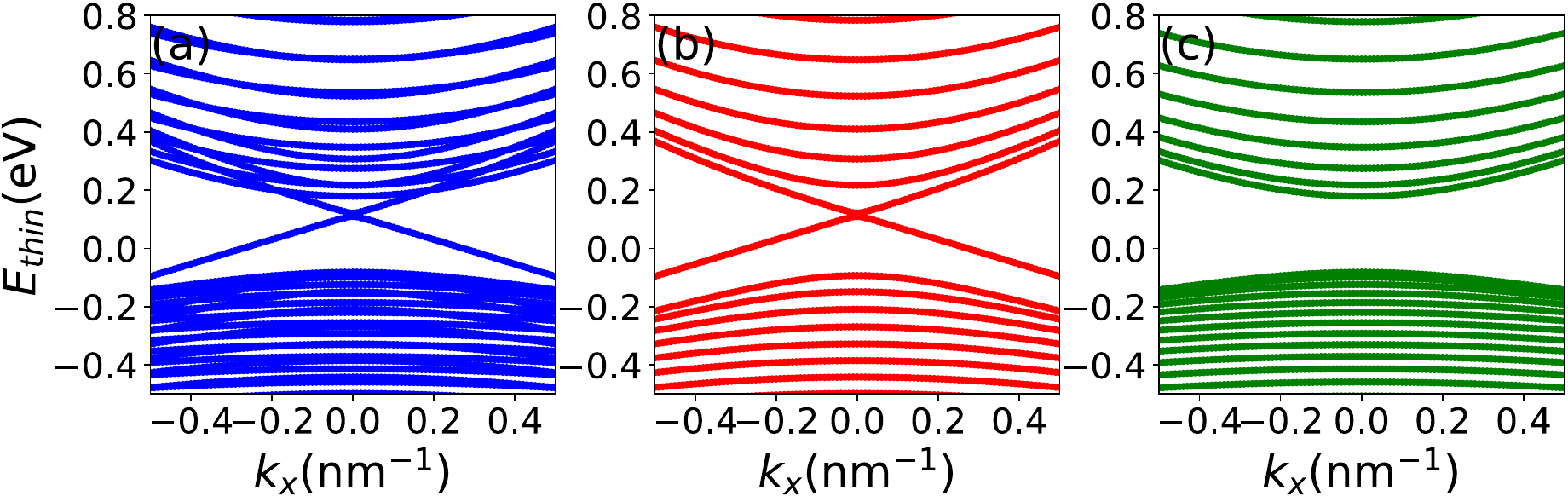}
\caption{Dispersions of the thin film for the Cr-doped Bi$_{2}$Se$_{3}$ under open boundary condition along $y$ direction and periodic boundary conditions along $x$ direction. (a) Bands for the Floquet Hamiltonian. (b) Bands for the pseudo-spin-$\uparrow$ (``+'') Floquet Hamiltonian. (c) Bands for the pseudo-spin-$\downarrow$ (``-'') Floquet Hamiltonian.
The other parameters are given as $A_{0}=0.036$ \AA$^{-1}$, $L=3$ nm, $a_x=a_y=1$ \AA, $N_y=200$, $m_{0}=30$ meV, $\varphi=\pi/2$, and $\hbar\omega\approx0.15$ eV ($\omega \sim 2.2789\times10^{2}$ THz) for chromium-doped Bi$_{2}$Se$_{3}$. Here, $a_x$ and $a_y$ are the lattice constants along $x$ and $y$ directions respectively.} \label{Fig:E_thinfilm}
\end{figure}

Furthermore, we calculate the energy dispersions of the thin film for the Cr-doped Bi$_{2}$Se$_{3}$ with $A_{0}=0.036$ \AA$^{-1}$ and $L=3$ nm under open boundary condition along $y$ direction and periodic boundary conditions along $x$ direction (the detail derivations can be found in the Appendix B).
As shown in Fig.~\ref{Fig:E_thinfilm}, it is found that there exist the edge states in the thin film as shown in Fig.~\ref{Fig:E_thinfilm}(a), and the edge states come from the pseudo-spin-$\uparrow$ (``$+$'') Floquet Hamiltonian in the chosen parameters here as shown in Fig.~\ref{Fig:E_thinfilm}(b). However, there is no edge states in the energy spectra of the pseudo-spin-$\downarrow$ (``$-$'') Floquet Hamiltonian as shown in Fig.~\ref{Fig:E_thinfilm}(c).

\section{Probation of the topological transitions}\label{8}

\begin{figure}[htpb]
\centering
\includegraphics[width=0.48\textwidth]{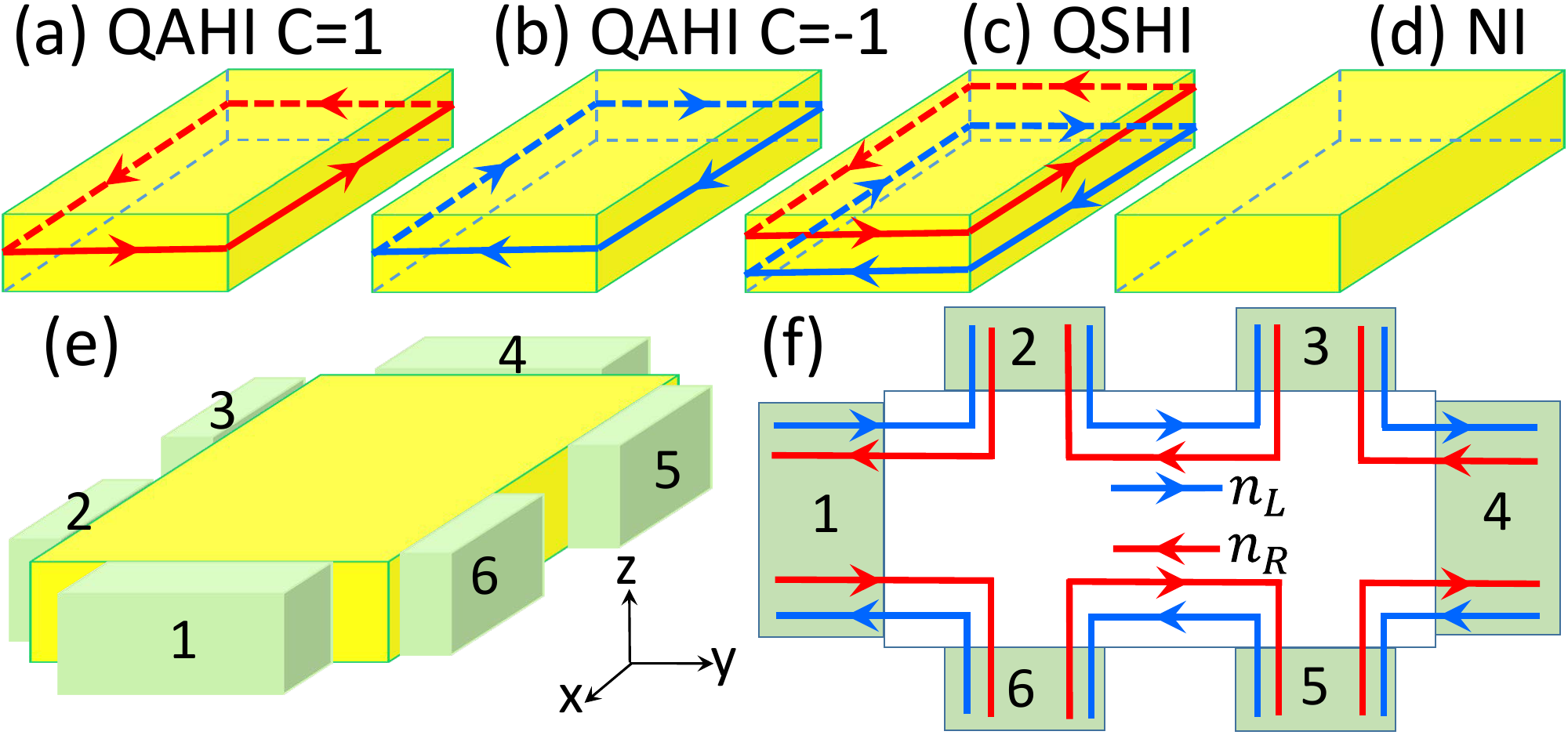}\\
\includegraphics[width=0.48\textwidth]{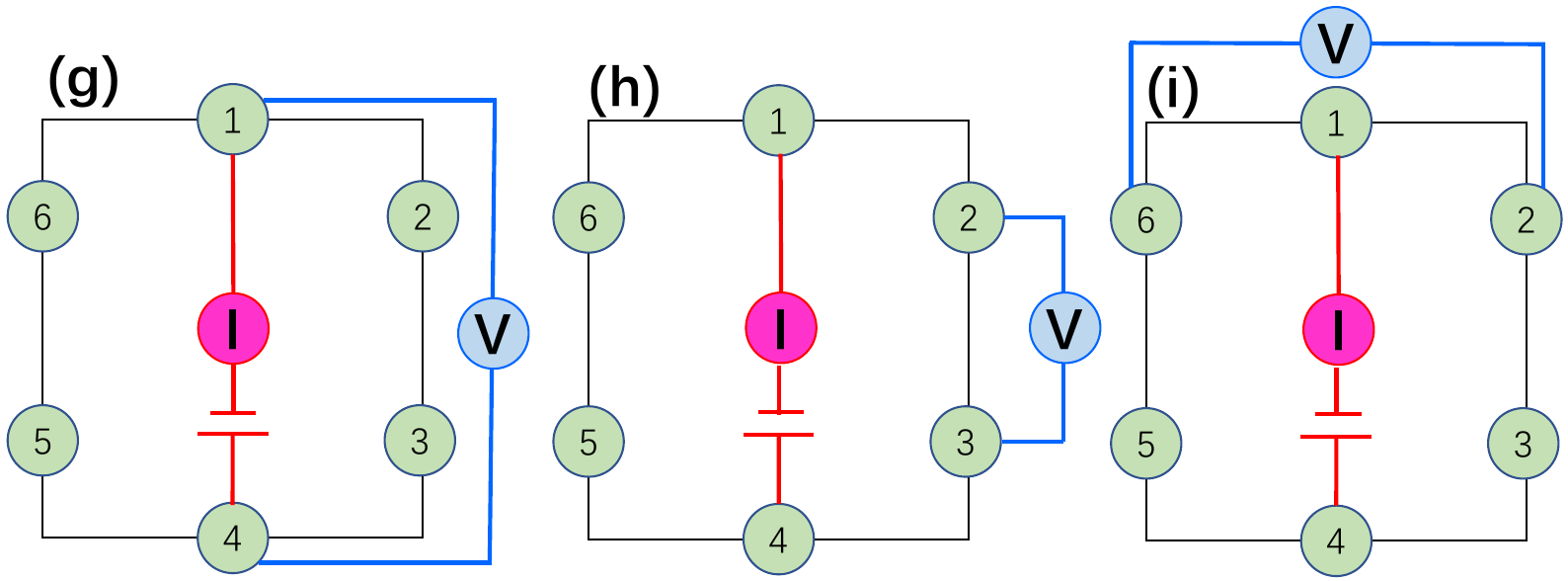}
\caption{
(a) QAHI with $C_{-}=0$ and $C_{+}=1$ which corresponds to pseudo-spin-$\uparrow$ (``$+$'') current on the side edges. Here, the arrow denotes the direction of the current.
(b) QAHI with $C_{+}=0$ and $C_{-}=-1$ which corresponds to pseudo-spin-$\downarrow$ (``$-$'') current on the side edges.
(c) QSHI has opposite currents with different pseudo-spins on the side edges.
(d) NI with no pseudo-spin currents. 
(e) The proposed device for nonlocal edge measurements with the electrodes (green cuboids).
(f) Schematics of the experimental setup with the electrodes (green rectangles). Here, $n_L$ and $n_R$ are the coefficients of the transmission probability matrix, and the number of the clockwise and anti-clockwise pseudo-spin currents are determined by $n_L$ and $n_R$.
(g)-(i) Electric-circuit diagrams with a bias voltage leads on electrodes 1 and 4, and the currents which leads on electrodes 2, 3, 5, and 6 are set to zero. In particular, the electric-circuit diagram (i) is used to measure the Hall resistance.} \label{Fig:nonlocal}
\end{figure}

Physically, as shown in Fig.~\ref{Fig:nonlocal}(a)-(d), the quantum anomalous Hall insulator with the Chern number $C=1$ has a pseudo-spin-$\uparrow$ (``$+$'') current on the side edges~\cite{XueExp2013,Qi2008,Yu2010,Chu2011,Qiao2010,Qi2006,Onoda2006,Nomura2011,Checkelsky2012,Hor2010,Chen2010,Wray2011,Wang2013,Wang2014,Law2017,Kim2018,Kawamura2018}, 
the quantum anomalous Hall insulator with the Chern number $C=-1$ has a pseudo-spin-$\downarrow$ (``$-$'') current on the side edges, and the quantum spin Hall insulator has antiparallel currents with different pseudo-spins on the side edges~\cite{Roth2009,Chen2021prb,Piskunow2021prl}. Here, each current is responsible for the $e^2/h$ Hall conductance as shown in Eq.~(\ref{eq:anomalousHallConductance}). However, the normal insulator has no pseudo-spin currents which correspond to a zero Hall conductance.

Detecting the pseudo-spin current can help to identify the topological transitions.
An experimental set up to measure the pseudo-spin current is given in Fig.~\ref{Fig:nonlocal}(d): The side surface of the device has six electrodes as 1-6 (green cuboids). Further, one needs to measure the nonlocal edge transport~\cite{Roth2009,Chen2021prb,Piskunow2021prl,Chen2012prb}, i.e., the nonlocal resistances. As shown in Fig.~\ref{Fig:nonlocal}(e) and (f), a current $I_{ij}$ is applied between electrodes $i$ and $j$ on the side surface, and a nonlocal voltage $V_{ij}$ between other two electrodes $i'$ and $j'$ is measured to define a nonlocal resistance $R_{ij,i'j'} = V_{i'j'}/I_{ij}$~\cite{Roth2009,Chen2021prb,Piskunow2021prl,Chen2012prb,Datta1997book}.

Within the Landauer-B{\"u}ttiker-Fisher-Lee formalism~\cite{Landauer1970pm,Buttiker1988prb,Buttiker1986prl,Fisher1981prb}, the current $I_{i}$ flowing out of the $i$th electrode into the sample region is expressed as~\cite{Roth2009,Chen2021prb,Piskunow2021prl,Chen2012prb,Datta1997book}
\begin{align}\label{eq:LBFL}
I_{i}=\frac{e^2}{h}\sum_{j,j\neq i}T_{ij}(E_F)(V_{i} - V_{j}),
\end{align}
where $V_i$ is the voltage on the $i$th electrode and $T_{ij}(E_F)$ is the transmission probability matrix from the $i$th to the $j$th electrode at Fermi energy $E_F$. Notice that, in a time-reversal-invariant system, the transmission
coefficients satisfy the condition $T_{ij}=T_{ji}$~\cite{Roth2009}.
For the present case here, there are six electrodes on the side surface, so $T_{ij}$ is a $6\times 6$ matrix. 
Concretely, Eq.~(\ref{eq:LBFL}) can be written in a matrix form
\begin{widetext}\begin{align}
\begin{pmatrix}
I_1 \\
I_2 \\
I_3 \\
I_4 \\
I_5 \\
I_6
\end{pmatrix}=\frac{e^2}{h}\begin{pmatrix}
\sum_{j=1}^{6}T_{1j} & -T_{12} & -T_{13} & -T_{14} & -T_{15} & -T_{16} \\
-T_{21} & \sum_{j=1}^{6}T_{2j} & -T_{23} & -T_{24} & -T_{25} & -T_{26} \\
-T_{31} & -T_{32} & \sum_{j=1}^{6}T_{3j} & -T_{34} & -T_{35} & -T_{36} \\
-T_{41} & -T_{42} & -T_{43} & \sum_{j=1}^{6}T_{4j} & -T_{45} & -T_{46} \\
-T_{51} & -T_{52} & -T_{53} & -T_{54} & \sum_{j=2}^{6}T_{5j} & -T_{56} \\
-T_{61} & -T_{62} & -T_{63} & -T_{64} & -T_{65} & \sum_{j=1}^{6}T_{6j}
\end{pmatrix}
\begin{pmatrix}
V_1 \\
V_2 \\
V_3 \\
V_4 \\
V_5 \\
V_6
\end{pmatrix},
\end{align}\end{widetext} where $T_{11}=T_{22}=T_{33}=T_{44}=T_{55}=T_{66}=0$.
If the six electrodes are identical, we have $T_{j+1,j}=T_{1,6}=n_R$ and $T_{j,j+1}=T_{6,1}=n_L$ with $j=1,2,\cdots,5$; $T_{j+2,j}=T_{2,6}=T_{1,5}$ and $T_{j,j+2}=T_{6,2}=T_{5,1}$ with $j=1,2,3,4$; $T_{j+3,j}=T_{3,6}$ and $T_{j,j+3}=T_{6,3}$, i.e., $T_{j+3,j}=T_{j,j+3}$ with $j=1,2,3$. For the non-neighboring terminals, we set that $T_{ij}=0$. Therefore, the corresponding transmission matrix has the form
\begin{align}
&T=\nonumber\\
&\!\begin{pmatrix}
n_L\!+\!n_R & \!-\!n_L & 0 & 0 & 0 & \!-\!n_R \\
\!-\!n_R & n_L\!+\!n_R & \!-\!n_L & 0 & 0 & 0 \\
0 & \!-\!n_R & n_L\!+\!n_R & \!-\!n_L & 0 & 0 \\
0 & 0 & \!-\!n_R & n_L\!+\!n_R & \!-\!n_L & 0 \\
0 & 0 & 0 & \!-\!n_R & n_L\!+\!n_R & \!-\!n_L \\
\!-\!n_L & 0 & 0 & 0 & \!-\!n_R & n_L\!+\!n_R
\end{pmatrix}.\!
\end{align}

As shown in Fig.~\ref{Fig:nonlocal}(g)-(i), we can apply a bias voltage leads on electrodes 1 and 4, and the currents which leads on electrodes 2, 3, 5, and 6 are set to zero. Owing to the current conservation, one finds that $I_1=-I_4\equiv I_{14}$. We set $V_5 = 0$, which allows us to truncate the fifth row and the fifth column of the matrix and write
\begin{align}\label{eq:Vs}
\begin{pmatrix}
V_1 \\
V_2 \\
V_3 \\
V_4 \\
V_6
\end{pmatrix}=\frac{h}{e^2}T_{d}^{-1}\begin{pmatrix}
I_{14} \\
0 \\
0 \\
-I_{14} \\
0
\end{pmatrix},
\end{align} where
\begin{align}
&T_{d}=\nonumber\\
&\!\begin{pmatrix}
n_L\!+\!n_R & \!-\!n_L & 0 & 0 & \!-\!n_R \\
\!-\!n_R & n_L\!+\!n_R & \!-\!n_L & 0 & 0 \\
0 & \!-\!n_R & n_L\!+\!n_R & \!-\!n_L & 0 \\
0 & 0 & \!-\!n_R & n_L\!+\!n_R & 0 \\
\!-\!n_L & 0 & 0 & 0 & n_L\!+\!n_R
\end{pmatrix}\!.
\end{align}
Then the voltages of the six terminals are given by Eq.~(\ref{eq:Vs}) as
\begin{align}
V_1&=\frac{hI_{14}}{e^2}\frac{n_R}{n_L^2+n_R^2-n_Ln_R},\\
V_2&=\frac{hI_{14}}{e^2}\frac{n_R^2+n_Ln_R-n_L^2}{n_L^3+n_R^3},\\
V_3&=\frac{hI_{14}}{e^2}\frac{n_R-n_L}{n_L^2+n_R^2-n_Ln_R},\\
V_4&=-\frac{hI_{14}}{e^2}\frac{n_L^2}{n_L^3+n_R^3},\\
V_5&=0,\\
V_6&=\frac{hI_{14}}{e^2}\frac{n_Ln_R}{n_L^3+n_R^3},
\end{align}
where $n_L$ and $n_R$ are the coefficients of the transmission probability matrix. They determine the number of the
clockwise and anti-clockwise pseudo-spin currents as shown in the Fig.~\ref{Fig:nonlocal}(f).
For the QAHI with $C=1$, we have $n_L = 1$ and $n_R = 0$. For the QAHI with $C=-1$, we have $n_L = 0$ and $n_R = 1$. For the QSHI, we have $n_L = n_R = 1$. For the NI, we have $n_L = n_R = 0$. As shown in Fig.~\ref{Fig:nonlocal}(g)-(i), the nonlocal resistances can be analytically found as
\begin{align}
R_{14,14}&=\frac{V_1 - V_4}{I_{14}}=\frac{h}{e^2}\frac{n_L^2 + n_R^2 + n_Rn_L}{n_L^3 + n_R^3},\\
R_{14,23}&=\frac{V_2 - V_3}{I_{14}}=\frac{h}{e^2}\frac{n_{L}n_{R}}{n_L^3 + n_R^3},\\
R_{14,26}&=\frac{V_2 - V_6}{I_{14}}=\frac{h}{e^2}\frac{n_R-n_L}{n_L^2 + n_R^2 - n_{L}n_{R}}.
\end{align}
For the QAHI with $C=1$, we have $R_{14,14} = \frac{h}{e^2}$, $R_{14,23} = 0$, and $R_{14,26} = -\frac{h}{e^2}$. For the QAHI with $C=-1$, we have $R_{14,14} = \frac{h}{e^2}$, $R_{14,23} = 0$,  and $R_{14,26} = \frac{h}{e^2}$. For the QSHI, we obtain $R_{14,14} = \frac{3h}{2e^2}$, $R_{14,23} = \frac{h}{2e^2}$, and $R_{14,26} = 0$. For the NI, we have $R_{14,14} = R_{14,23} = R_{14,26} = 0$.

\section{Summary}\label{9}

We investigate the impact of high-frequency pumping on a ferromagnetic topological insulator with the circularly polarized optical field. It is found that the intensity of the circularly polarized light can be used as a knob to drive a topological transition.
With modulating the strength of the polarized optical field in an experimentally accessible range, there are four different regions: the NI phase, the time-reversal-symmetry-broken QSHI phase, and two different QAHI phases. This is different from the situation in the absent of optical field.

We propose an experimental scheme to manipulate the topological phases in Cr-doped Bi$_{2}$Se$_{3}$ with high-frequency pumping light. Our proposal can be realized in an experimentally accessible range. Particularly, to realize the light driven topological phases, the frequency and intensity of the light are both within the experimental accessibility~\cite{lightexp2013,lightexp2016}. In most of the recent experiments~\cite{XueExp2013,CheckelskyExp2014,QAHIexp2014,QAHIexp2015,XueExp2015,YoshimiExp2015,ChangExp2016,GrauerExp2017,XueExp2018,ZhangExp2020,QAHIexp2020,ZhaoExp2020,MogiExp2021}, people focus on using the magnetic fields to manipulate the topological phases of the system.
However, the topological phases obtained in this way may be confused with the quantum Hall effect. Luckily, our proposal avoids this. Therefore, the theoretically investigations we put forward will be helpful to the future experiments.

Note Added. When finishing this paper, we became aware of another preprint, Ref.~\cite{Liu2021arxive}, which also did the Floquet engineering of magnetism in topological insulator thin films.

\begin{acknowledgments}
Fang Qin and Rui Chen contribute equally to this work. We acknowledge helpful discussions with Dong-Hui Xu, Hai-Peng Sun, Xiao-Bin Qiang, De-Huan Cai, and Hui Xiong.
This work was supported by the National Natural Science Foundation of China (Grants No.~11534001, No.~11925402, and No.~11404106), the Strategic Priority Research Program of Chinese Academy of Sciences (Grant No.~XDB28000000), Guangdong province (Grants No.~2016ZT06D348 and No.~2020KCXTD001), the National Key R \& D Program (Grant No.~2016YFA0301700), Shenzhen High-level Special Fund (Grants No.~G02206304 and No.~G02206404), and the Science, Technology and Innovation Commission of Shenzhen Municipality (No.~ZDSYS20190902092905285, No.~ZDSYS20170303165926217, No.~JCYJ20170412152620376, and No.~KYTDPT20181011104202253).
F.Q. acknowledges support from the project funded by the China Postdoctoral Science Foundation (Grant No.~2019M662150 and No.~2020T130635) and the SUSTech Presidential Postdoctoral Fellowship.
R.C. acknowledges support from the project funded by the China Postdoctoral Science Foundation (Grant No.~2019M661678) and the SUSTech Presidential Postdoctoral Fellowship.
\end{acknowledgments}

\appendix
\section{Derivations of Eq.~(\ref{eq:E0zf})}

\subsection{Semi-infinite boundary conditions}\label{sec:Semi-infinite-thickness}

The surface states have a finite distribution near the boundary. For a film thick enough that the
states at opposite surfaces barely couple to each other, we can focus on just one surface. Without
loss of generality, we study a system from $z = 0$ to $+\infty$. The boundary condition is given as
\begin{align}\label{eq: semi-infinite-thickness}
\Psi_{\uparrow}(z=0) = 0,~~~~~~~\Psi_{\uparrow}(z=+\infty) = 0.
\end{align}
The condition of $\Psi_{\uparrow}(z=+\infty) = 0$ requires that $\Psi_{\uparrow}$ contains only the four terms in which $\beta$ is
negative and that the real part of $\lambda_{\alpha}$ be positive.

With Eq.~(\ref{eq:wavefunctionup1}), we have 
\begin{align}\label{eq:wavefunctionup-semi-infinite01}
\Psi_{\uparrow}(z) &= C_{1-} \left(
  \begin{array}{cc}
    D_{+}\lambda_{1}^{2} - L_{-} + E_{0}  \\
    i A_{1}\lambda_{1}
  \end{array}
\right)  e^{-\lambda_{1}z} \nonumber\\
&~~+  C_{2-} \left(
  \begin{array}{cc}
    D_{+}\lambda_{2}^{2} - L_{-} + E_{0}  \\
   i A_{1}\lambda_{2}
  \end{array}
\right)  e^{-\lambda_{2}z} .
\end{align} 

Applying the boundary conditions of equation (\ref{eq: semi-infinite-thickness}) to the general solution of equation (\ref{eq:wavefunctionup-semi-infinite01}),
the secular equation of the nontrivial solution to the coefficients $C_{\alpha\beta}$ leads to
\begin{align}
( E_{0} \!-\! L_{-})(C_{1-} \!+\! C_{2-} ) + D_{+}(C_{1-} \lambda_{1}^{2} \!+\! C_{2-} \lambda_{2}^{2}) &\!=\!0, \label{eq:semi-infinite-1}\\
C_{1-}\lambda_{1} \!+\! C_{2-}\lambda_{2}&\!=\!0.
\end{align} 
Substituting $C_{2-}=-C_{1-}\lambda_{1}/\lambda_{2}$ into Eq.~(\ref{eq:semi-infinite-1}), we have
\begin{align}
E_{0} - L_{-} &= D_{+}\lambda_{1}\lambda_{2} = \frac{\sqrt{F^{2} - R}}{2D_{-}} \nonumber\\
&= \frac{\sqrt{D_{+} D_{-}(E_{0} - L_{+}) (E_{0} - L_{-})}}{D_{-}}.
\end{align} 
Therefore, we have
\begin{align}
E_{\pm}^{0}&\equiv E_{0}= \frac{D_{+}L_{+} - D_{-}L_{-} }{D_{+} - D_{-} } = \frac{D_{1}M_{0} + B_{1}C_{0} }{B_{1} } \nonumber\\
&= C_{0} + \frac{D_{1}M_{0} }{B_{1} }.
\end{align}

\section{Tight-binding model under open boundary condition along $y$ direction and periodic boundary conditions along $x$ direction}\label{4}

In a lattice, one makes the following replacements~\cite{Shen2012}
\begin{align}
&k_{j}\rightarrow\frac{1}{a_j}\sin(k_{j}a_{j}),\\
&k_{j}^{2}\rightarrow\frac{2}{a_j^2}[1-\cos(k_{j}a_{j})],
\end{align} where $j=x,y,z$, $a_j$ is the lattice constant along $j$ direction, $\sin(k_{j}a_{j})=\frac{e^{ik_{j}a_{j}}-e^{-ik_{j}a_{j}}}{2i}$ and $\cos(k_{j}a_{j})=\frac{e^{ik_{j}a_{j}}+e^{-ik_{j}a_{j}}}{2}$.
\begin{align}
&k_{\perp}^{2}=k_{x}^{2}+k_{y}^{2}\rightarrow\frac{2}{a_x^2}[1-\cos(k_{x}a_{x})] + \frac{2}{a_y^2}[1-\cos(k_{y}a_{y})].
\end{align}
In this way the hopping terms in the lattice model only exist between the nearest neighbor sites.
With this mapping, one obtains the following tight-binding model for topological thin film with high-frequency pumping in the basis $(c_{{\bf k},+,\uparrow}, c_{{\bf k},-,\uparrow}, c_{{\bf k},+,\downarrow}, c_{{\bf k},-,\downarrow})^{T}$ with ${\bf k}=(k_x,k_y)$ as 
\begin{align}\label{eq:HFfilm_supp}
\!H_{\text{film}}^{(F)}\!=\! \left(
  \begin{array}{cccc}
    h_{f1}(k_{x},k_y) & 0  \\
    0 & h_{f2}(k_{x},k_y)
  \end{array}
\right),
\end{align}
where
\begin{align}
&~~\!h_{f1}(k_x,k_y) \nonumber\\
&\!=\! \left(\tilde{E}_{1}^{0} \!-\! D\left\{\frac{2}{a_x^2}[1\!-\!\cos(k_{x}a_{x})] \!+\! \frac{2}{a_y^2}[1\!-\!\cos(k_{y}a_{y})] \right\} \right)\sigma_0  \nonumber\\
&\!+\! \left(\frac{\tilde{\Delta}}{2} \!-\! B\left\{\frac{2}{a_x^2}[1\!-\!\cos(k_{x}a_{x})] \!+\! \frac{2}{a_y^2}[1\!-\!\cos(k_{y}a_{y})] \right\}\!+\! \tilde{m} \right)\sigma_z \nonumber\\
&\!+\! i\gamma_{1} \left(\frac{1}{a_x}\sin(k_{x}a_{x}) \!-\! \frac{i}{a_y}\sin(k_{y}a_{y}) \right)\sigma_{+}\nonumber\\
&\!-\! i\gamma_{1} \left(\frac{1}{a_x}\sin(k_{x}a_{x}) \!+\! \frac{i}{a_y}\sin(k_{y}a_{y}) \right)\sigma_{-}, 
\end{align}
\begin{align}
&~~\!h_{f2}(k_x,k_y) \nonumber\\
&\!=\! \left(\tilde{E}_{2}^{0} \!-\! D\left\{\frac{2}{a_x^2}[1\!-\!\cos(k_{x}a_{x})] \!+\! \frac{2}{a_y^2}[1\!-\!\cos(k_{y}a_{y})] \right\} \right)\sigma_0  \nonumber\\
&\!+\! \left(-\frac{\tilde{\Delta}}{2} \!+\! B\left\{\frac{2}{a_x^2}[1\!-\!\cos(k_{x}a_{x})] \!+\! \frac{2}{a_y^2}[1\!-\!\cos(k_{y}a_{y})] \right\}\!+\! \tilde{m} \right)\sigma_z \nonumber\\
&\!+\! i\gamma_{2} \left(\frac{1}{a_x}\sin(k_{x}a_{x}) \!-\! \frac{i}{a_y}\sin(k_{y}a_{y}) \right)\sigma_{+}\nonumber\\
&\!-\! i\gamma_{2} \left(\frac{1}{a_x}\sin(k_{x}a_{x}) \!+\! \frac{i}{a_y}\sin(k_{y}a_{y}) \right)\sigma_{-}.
\end{align}

Performing the Fourier transformation, one obtains
\begin{align}
c_{{\bf k},s,\sigma}&=\frac{1}{\sqrt{N_{y}}}\sum_{k_x,j_y}^{N_y}e^{-ik_{y}j_{y}a_{y}}c_{k_x,j_y,s,\sigma},\\
c_{{\bf k},s,\sigma}^{\dagger}&=\frac{1}{\sqrt{N_{y}}}\sum_{k_x,j_y}^{N_y}e^{ik_{y}j_{y}a_{y}}c_{k_x,j_y,s,\sigma}^{\dagger}.
\end{align}
\begin{align}
C_{{\bf k},\sigma}&=\begin{pmatrix}
c_{{\bf k},+,\sigma} \\
c_{{\bf k},-,\sigma}
\end{pmatrix}=\frac{1}{\sqrt{N_{y}}}\sum_{k_x,j_y}^{N_y}e^{-ik_{y}j_{y}a_{y}}C_{k_x,j_y,\sigma},\\
C_{{\bf k},\sigma}^{\dagger}&=\begin{pmatrix}
c_{{\bf k},+,\sigma}^{\dagger} &
c_{{\bf k},-,\sigma}^{\dagger}
\end{pmatrix}=\frac{1}{\sqrt{N_{y}}}\sum_{k_x,j_y}^{N_y}e^{ik_{y}j_{y}a_{y}}C_{k_x,j_y,\sigma}^{\dagger}.
\end{align} 

\begin{align}
&~~\sum_{{\bf k},s}C_{{\bf k},\uparrow}^{\dagger}\left[h_{f1}(k_{x},k_y)\right]C_{{\bf k},\uparrow}\nonumber\\ 
&\!=\!\sum_{k_x,j_y}\!\left[\!\tilde{E}_{1}^{0}\!-\!\frac{2D}{a_x^2}\!-\!\frac{2D}{a_y^2}\!+\!\frac{2D}{a_x^2}\cos(k_xa_x)\!\right]\!C_{k_x,j_y,\uparrow}^{\dagger}\sigma_{0}C_{k_x,j_y,\uparrow} \nonumber\\ 
&\!+\! \frac{D}{a_y^2}\sum_{k_x,j_y}\!\left[\!C_{k_x,j_y,\uparrow}^{\dagger}\sigma_{0}C_{k_x,j_y+1,\uparrow}\!+\! {\rm h.c.}\!\right]\! \nonumber\\
&\!+\!\sum_{k_x,j_y}\!\left[\!\frac{\tilde{\Delta}}{2}\!-\!\frac{2B}{a_x^2}\!-\!\frac{2B}{a_y^2}\!+\!\frac{2B}{a_x^2}\cos(k_xa_x)\!+\!\tilde{m}\!\right]\!\nonumber\\ 
&\!\times\!C_{k_x,j_y,\uparrow}^{\dagger}\sigma_{z}C_{k_x,j_y,\uparrow} \nonumber\\ 
&\!+\! \frac{B}{a_y^2}\sum_{k_x,j_y}\!\left[\!C_{k_x,j_y,\uparrow}^{\dagger}\!\sigma_{z}\!C_{k_x,j_y+1,\uparrow}\!+\! {\rm h.c.}\!\right]\! \nonumber\\
&\!+\!\frac{i\gamma_1}{a_x}\sum_{k_x,j_y}\sin(k_xa_x)\!\left[\!C_{k_x,j_y,\uparrow}^{\dagger}(\sigma_{+}-\sigma_{-})C_{k_x,j_y,\uparrow} \!\right]\! \nonumber\\
&\!-\!\frac{\gamma_1}{2a_y}\sum_{k_x,j_y}\!\left[\!iC_{k_x,j_y,\uparrow}^{\dagger}(\sigma_{+}+\sigma_{-})C_{k_x,j_y+1,\uparrow} \!+\! {\rm h.c.}\!\right].
\end{align}

\begin{align}
&~~\sum_{{\bf k},s}C_{{\bf k},\downarrow}^{\dagger}\left[h_{f2}(k_{x},k_y)\right]C_{{\bf k},\downarrow}\nonumber\\ 
&\!=\!\sum_{k_x,j_y}\!\left[\!\tilde{E}_{2}^{0}\!-\!\frac{2D}{a_x^2}\!-\!\frac{2D}{a_y^2}\!+\!\frac{2D}{a_x^2}\cos(k_xa_x)\!\right]\!C_{k_x,j_y,\downarrow}^{\dagger}\sigma_{0}C_{k_x,j_y,\downarrow} \nonumber\\ 
&\!+\! \frac{D}{a_y^2}\sum_{k_x,j_y}\!\left[\!C_{k_x,j_y,\uparrow}^{\dagger}\sigma_{0}C_{k_x,j_y+1,\downarrow}\!+\! {\rm h.c.}\!\right]\! \nonumber\\
&\!+\!\sum_{k_x,j_y}\!\left[\!-\!\frac{\tilde{\Delta}}{2}\!+\!\frac{2B}{a_x^2}\!+\!\frac{2B}{a_y^2}\!-\!\frac{2B}{a_x^2}\cos(k_xa_x)\!+\!\tilde{m}\!\right]\!\nonumber\\ 
&\!\times\!C_{k_x,j_y,\downarrow}^{\dagger}\sigma_{z}C_{k_x,j_y,\downarrow} \nonumber\\ 
&\!-\! \frac{B}{a_y^2}\sum_{k_x,j_y}\!\left[\!C_{k_x,j_y,\downarrow}^{\dagger}\sigma_{z}C_{k_x,j_y+1,\downarrow}\!+\! {\rm h.c.}\!\right]\! \nonumber\\
&\!+\!\frac{i\gamma_2}{a_x}\sum_{k_x,j_y}\sin(k_xa_x)\!\left[\!C_{k_x,j_y,\downarrow}^{\dagger}(\sigma_{+}-\sigma_{-})C_{k_x,j_y,\downarrow} \!\right]\! \nonumber\\
&\!-\!\frac{\gamma_2}{2a_y}\sum_{k_x,j_y}\!\left[\!iC_{k_x,j_y,\downarrow}^{\dagger}(\sigma_{+}+\sigma_{-})C_{k_x,j_y+1,\downarrow} \!+\! {\rm h.c.}\!\right].
\end{align}

Therefore, the tight-binding Hamiltonian under $x$-PBCs and $y$-OBCs in the basis $(C_{k_x,1,\uparrow}, C_{k_x,1,\downarrow}, C_{k_x,2,\uparrow}, C_{k_x,2,\downarrow},\cdots, C_{k_x,N_y,\uparrow}, C_{k_x,N_y,\downarrow})^{T}$ is given by 
\begin{align}
H_{y-OBC}=\begin{pmatrix}
h & T & 0 & \cdots & 0 \\
T^{\dagger} & h & T & \cdots & 0 \\
0 & T^{\dagger} & h & \ddots & \vdots \\
\vdots & \ddots & \ddots & \ddots & T \\
0& \cdots & 0 & T^{\dagger} & h
\end{pmatrix}_{(4N_y)\times (4N_y)},
\end{align} where
\begin{align}
&h\!=\!\begin{pmatrix}\!
\bar{E}_{1}^{0}\sigma_0 \!+\! \bar{\Delta}_{1}\sigma_z \!+\! i\bar{\gamma}_1(\!\sigma_{+}\!-\!\sigma_{-}\!) & 0 \\
0 & \bar{E}_{2}^{0}\sigma_0 \!+\! \bar{\Delta}_{2}\sigma_z \!+\! i\bar{\gamma}_2(\!\sigma_{+}\!-\!\sigma_{-}\!)
\!\end{pmatrix},\\
&\bar{E}_{1}^{0}=\tilde{E}_{1}^{0}\!-\!\frac{2D}{a_x^2}\!-\!\frac{2D}{a_y^2}\!+\!\frac{2D}{a_x^2}\cos(k_xa_x), \\
&\bar{E}_{2}^{0}=\tilde{E}_{2}^{0}\!-\!\frac{2D}{a_x^2}\!-\!\frac{2D}{a_y^2}\!+\!\frac{2D}{a_x^2}\cos(k_xa_x), \\
&\bar{\Delta}_{1}=\frac{\tilde{\Delta}}{2}\!-\!\frac{2B}{a_x^2}\!-\!\frac{2B}{a_y^2}\!+\!\frac{2B}{a_x^2}\cos(k_xa_x)\!+\!\tilde{m},\\
&\bar{\Delta}_{2}=-\frac{\tilde{\Delta}}{2}\!+\!\frac{2B}{a_x^2}\!+\!\frac{2B}{a_y^2}\!-\!\frac{2B}{a_x^2}\cos(k_xa_x)\!+\!\tilde{m},\\
&\bar{\gamma}_1=\frac{\gamma_1}{a_x}\sin(k_xa_x),\\
&\bar{\gamma}_2=\frac{\gamma_2}{a_x}\sin(k_xa_x),
\end{align}
\begin{align}
&T\!=\!\begin{pmatrix}\!
\frac{D}{a_y^2}\sigma_0 \!+\! \frac{B}{a_y^2}\sigma_z \!-\! \frac{i\gamma_1}{2a_y}\sigma_{x} & 0 \\
0 & \frac{D}{a_y^2}\sigma_0 \!-\! \frac{B}{a_y^2}\sigma_z \!-\! \frac{i\gamma_2}{2a_y}\sigma_{x}
\!\end{pmatrix}\!,\\
&T^{\dagger}\!=\!\begin{pmatrix}\!
\frac{D}{a_y^2}\sigma_0 \!+\! \frac{B}{a_y^2}\sigma_z \!+\! \frac{i\gamma_1}{2a_y}\sigma_{x} & 0 \\
0 & \frac{D}{a_y^2}\sigma_0 \!-\! \frac{B}{a_y^2}\sigma_z \!+\! \frac{i\gamma_2}{2a_y}\sigma_{x}
\!\end{pmatrix}\!,
\end{align} where we use $\sigma_{+}\!+\!\sigma_{-}=\sigma_{x}$.

In addition, we can also get the tight-binding Hamiltonian with pseudo-spin-$\uparrow$ (``+'') under $x$-PBCs and $y$-OBCs in the basis $(C_{k_x,1,\uparrow}, C_{k_x,2,\uparrow},\cdots,C_{k_x,N_y,\uparrow})^{T}$ as
\begin{align}
H_{y-OBC}^{\uparrow}=\begin{pmatrix}
h_{\uparrow} & T_{\uparrow} & 0 & \cdots & 0 \\
T_{\uparrow}^{\dagger} & h_{\uparrow} & T_{\uparrow} & \cdots & 0 \\
0 & T_{\uparrow}^{\dagger} & h_{\uparrow} & \ddots & \vdots \\
\vdots & \ddots & \ddots & \ddots & T_{\uparrow} \\
0& \cdots & 0 & T_{\uparrow}^{\dagger} & h_{\uparrow}
\end{pmatrix}_{(2N_y)\times (2N_y)},
\end{align} where
\begin{align}
h_{\uparrow}\!=\!
\bar{E}_{1}^{0}\sigma_0 \!+\! \bar{\Delta}_{1}\sigma_z \!+\! i\bar{\gamma}_1(\!\sigma_{+}\!-\!\sigma_{-}\!),
\end{align}
\begin{align}
&T_{\uparrow}\!=\!
\frac{D}{a_y^2}\sigma_0 \!+\! \frac{B}{a_y^2}\sigma_z \!-\! \frac{i\gamma_1}{2a_y}\sigma_{x},\\
&T^{\dagger}_{\uparrow}\!=\!
\frac{D}{a_y^2}\sigma_0 \!+\! \frac{B}{a_y^2}\sigma_z \!+\! \frac{i\gamma_1}{2a_y}\sigma_{x}.
\end{align}

Further, we can get the tight-binding Hamiltonian with pseudo-spin-$\downarrow$ (``-'') under $x$-PBCs and $y$-OBCs in the basis $(C_{k_x,1,\downarrow}, C_{k_x,2,\downarrow},\cdots, C_{k_x,N_y,\downarrow})^{T}$ as
\begin{align}
H_{y-OBC}^{\downarrow}=\begin{pmatrix}
h_{\downarrow} & T_{\downarrow} & 0 & \cdots & 0 \\
T_{\downarrow}^{\dagger} & h_{\downarrow} & T_{\downarrow} & \cdots & 0 \\
0 & T_{\downarrow}^{\dagger} & h_{\downarrow} & \ddots & \vdots \\
\vdots & \ddots & \ddots & \ddots & T_{\downarrow} \\
0& \cdots & 0 & T_{\downarrow}^{\dagger} & h_{\downarrow}
\end{pmatrix}_{(2N_y)\times (2N_y)},
\end{align} where
\begin{align}
h_{\downarrow}\!=\!
\bar{E}_{2}^{0}\sigma_0 \!+\! \bar{\Delta}_{2}\sigma_z \!+\! i\bar{\gamma}_2(\!\sigma_{+}\!-\!\sigma_{-}\!),
\end{align}
\begin{align}
&T_{\downarrow}\!=\!
\frac{D}{a_y^2}\sigma_0 \!-\! \frac{B}{a_y^2}\sigma_z \!-\! \frac{i\gamma_2}{2a_y}\sigma_{x},\\
&T^{\dagger}_{\downarrow}\!=\!
\frac{D}{a_y^2}\sigma_0 \!-\! \frac{B}{a_y^2}\sigma_z \!+\! \frac{i\gamma_2}{2a_y}\sigma_{x}.
\end{align}

\end{document}